\documentclass[a4paper,12pt, epsfig]{article}
\pdfoutput=1
\usepackage{epsfig}
\usepackage{epstopdf}
\usepackage{graphicx}
\usepackage{ifthen}
\usepackage{feynmp}
\usepackage{amsmath}
\usepackage[psamsfonts]{amssymb}
\usepackage{euscript}

\usepackage{latexsym}
\usepackage[arrow,matrix,curve]{xy}

\newskip\humongous \humongous=0pt plus 1000pt minus 1000pt

\newif\ifdtup

\def\,{\hspace{-.1cm}}
\def\hsp{,\hspace{.7cm}}

\def\tf {\tilde{f}}

\def\fc#1#2 {\frac{n}{q}#1\frac{n}{q}#2}

\def\bp{\mathbf{P}}

\def\Os{|\Omega\rangle}
\def\mo#1{\int\frac{d{#1}}{2\pi}}
\renewcommand{\cos}{\textrm{cos}}
\renewcommand{\sin}{\textrm{sin}}
\newcommand{\asin}{\textrm{arcsin}}
\renewcommand{\sinh}{\textrm{sinh}}
\renewcommand{\cosh}{\textrm{cosh}}
\renewcommand{\tanh}{\textrm{tanh}}
\newcommand{\sech}{\textrm{sech}}
\newcommand{\csch}{\textrm{csch}}

\def\exp#1{\hbox{\rm exp}\left(#1\right)}

\renewcommand{\theequation}{\arabic{section}.\arabic{equation}}
\renewcommand{\(}{\begin{equation}}
\renewcommand{\)}{end{equation} \vspace{-.05in}\linebreak}

\newcounter{saveeqn}
\newcounter{savealpheqn}

\newcommand{\alpheqn}{\setcounter{saveeqn}{\value{equation}}%
  \stepcounter{saveeqn}\setcounter{equation}{0}%
  \renewcommand{\theequation}{\mbox{\arabic{section}.\arabic{saveeqn}
\alph{equation}}}
  \renewcommand{\)}{\end{equation}}}
\def\part#1{\frac{\partial}{\partial{#1}}}%
\def\group#1{\refstepcounter{equation}\setcounter{saveeqn}
 {\value{equation}}%
  \label{#1}\setcounter{equation}{0}%
\renewcommand{\theequation}{\mbox{\arabic{section}.\arabic{saveeqn}
\alph{equation}}}
  \renewcommand{\)}{\end{equation}}}
\newcommand{\reseteqn}{\setcounter{equation}{\value{saveeqn}}%
  \renewcommand{\theequation}{\arabic{section}.\arabic{equation}}%
  \renewcommand{\)}{\end{equation}}}

\newcommand{\aalpheqn}{\setcounter{saveeqn}{\value{equation}}%
  \stepcounter{saveeqn}\setcounter{equation}{0}%
  \renewcommand{\theequation}{\mbox{
        \Alph{subsection}.\arabic{saveeqn}\alph{equation}}}
   \renewcommand{\)}{\end{equation}}}
\newcommand{\areseteqn}{\setcounter{equation}{\value{saveeqn}}%
  \renewcommand{\theequation}{\Alph{subsection}.\arabic{equation}}%
  \renewcommand{\)}{\end{equation}}}

\renewcommand{\thefootnote}{\alph{footnote}}
\renewcommand{\(}{\begin{equation}}
\renewcommand{\)}{\end{equation}}
\newcommand{\ba}{\begin{eqnarray}}
\newcommand{\ea}{\end{eqnarray}}
\newcommand{\cbp}{\mathop{\vtop{\ialign{##\crcr
   $\hfil\displaystyle{}\hfil$\crcr\noalign{\kern-13pt\nointerlineskip}
   \BIG{)}\hskip0pt\crcr\noalign{\kern3pt}}}}}
\newcommand{\pa}{\mathop{\vtop{\ialign{##\crcr
    
$\hfil\displaystyle{\oplus}\hfil$\crcr\noalign{\kern+1pt\nointerlineskip 
}
   \hspace{.08in}$^{\alpha=0}$\hskip6pt\crcr\noalign{\kern3pt}}}}}
\renewcommand{\hsp}{,\hspace{.3in}}
\newcommand{\p}{^\prime}

\catcode`\@=11
\def\vereq#1#2{\lower3pt\vbox{\baselineskip1.5pt \lineskip1.5pt
\ialign{$\m@th#1\hfill##\hfil$\crcr#2\crcr\sim\crcr}}}
\catcode`\@=12

\renewcommand{\(}{\begin{equation}}
\renewcommand{\)}{\end{equation}}

\def\pin#1{\int \frac{d#1}{2\pi}}

\def\df{\mathcal{D}_f}

\newcommand{\beas}{\begin{eqnarray*}}
\newcommand{\eeas}{\end{eqnarray*}}

\newcommand{\bquo}{\begin{quote}}
\newcommand{\enqu}{\end{quote}}

\def\bp{{\bf{p}}}
\def\bq{{\bf{q}}}
\def\bk{{\bf{k}}}
\def\ch{{\mathcal{H}}}
\def\co{{\mathcal{O}}}
\def\tf{{\tilde{f}}}

\newcommand{\beq}{\begin{equation}}
\newcommand{\eeq}{\end{equation}}
\newcommand{\bea}{\begin{eqnarray}}
\newcommand{\eea}{\end{eqnarray}}

\newskip\humongous \humongous=0pt plus 1000pt minus 1000pt

\newif\ifdtup

\catcode`\@=11

\@addtoreset{equation}{section}

\def\@normalsize{\@setsize\normalsize{15pt}\xiipt\@xiipt
\abovedisplayskip 14pt plus3pt minus3pt%
\belowdisplayskip \abovedisplayskip
\abovedisplayshortskip \z@ plus3pt%
\belowdisplayshortskip 7pt plus3.5pt minus0pt}

\def\small{\@setsize\small{13.6pt}\xipt\@xipt
\abovedisplayskip 13pt plus3pt minus3pt%
\belowdisplayskip \abovedisplayskip
\abovedisplayshortskip \z@ plus3pt%
\belowdisplayshortskip 7pt plus3.5pt minus0pt
\def\@listi{\parsep 4.5pt plus 2pt minus 1pt
      \itemsep \parsep
      \topsep 9pt plus 3pt minus 3pt}}

\relax

\catcode`@=12

\catcode`\@=11

\def\section{\@startsection{section}{1}{\z@}{3.5ex plus 1ex minus  .2ex}{2.3ex plus .2ex}{\large\bf}}

\def\thesection{\arabic{section}}
\def\thesubsection{\arabic{section}.\arabic{subsection}}

\def\appendix{\setcounter{section}{0}
 \def\thesection{Appendix \Alph{section}}
 \def\thesubsection{\Alph{section}.\arabic{subsection}}
 \def\theequation{\Alph{section}.\arabic{equation}}}
\renewcommand{\theequation}{\arabic{section}.\arabic{equation}}

\begin{document}
\def\thefootnote{\fnsymbol{footnote}}
\def\thetitle{Manifestly Finite Derivation of the Quantum Kink Mass}
\def\autone{Jarah Evslin}
\def\affa{Institute of Modern Physics, NanChangLu 509, Lanzhou 730000, China}
\def\affb{University of the Chinese Academy of Sciences, YuQuanLu 19A, Beijing 100049, China}

\begin{center}
{\large {\bf \thetitle}}

\bigskip

\bigskip

{\large \noindent  \autone{${}^{1,2}$}\footnote{jarah@impcas.ac.cn}}

\vskip.7cm

1) \affa\\
2) \affb\\

\end{center}

\begin{abstract}
\noindent
In 1974 Dashen, Hasslacher and Neveu calculated the leading quantum correction to the mass of the kink in the scalar $\phi^4$ theory in 1+1 dimensions.   The derivation relies on the identification of the perturbations about the kink as solutions of the P\"oschl-Teller (PT) theory.  They regularize the theory by placing it in a periodic box, although the kink is not itself periodic.  They also require an {\it{ad hoc}} identification of plane wave and PT states which is difficult to interpret in the decompactified limit.  We rederive the mass using the kink operator to recast this problem in terms of the PT Hamiltonian which we explicitly diagonalize using its exact eigenstates.  We normal order from the beginning, rendering our theory finite so that no compactification is necessary.    In our final expression for the kink mass, the form of the PT potential disappears, suggesting that our mass formula applies to other quantum solitons.

\end{abstract}

%
\setcounter{footnote}{0}
\renewcommand{\thefootnote}{\arabic{footnote}}

\section{Introduction}

In quantum field theory, particles are created by the creation operators $a^\dag$.  In contrast, solitons seem to be very different objects, corresponding somehow to classical solutions which must be quantum corrected \cite{gervais}.  This treatment of solitons is sufficient in many weakly coupled theories, but in theories which are strongly coupled in the infrared, it is no longer clear that classical solutions will be realized in the quantum theory.  Conversely, in strongly coupled theories, objects may be realized in the quantum theory which do not correspond to classical solutions\footnote{In fact, even at weak but finite coupling Derrick's theorem may be violated by quantum corrections \cite{delfino,davies} and so unstable classical configurations may correspond to stable quantum states.}  but nonetheless carry the same charges.  Therefore it would be desirable to have a description of these quantum objects which may be decoupled from classical solutions.

There is such a description.  Solitons in a quantum field theory are also represented by operators which create them, just like particles.   In the case of classical solutions, these are the operators which create coherent states \cite{hepp}.  Perhaps the first example of such an operator was the quantum kink, described by Mandelstam in Ref.~\cite{mandelstamkink}.  This kink leads to expectation values of the scalar field which reproduce the asymptotic behavior of the kink solution.  However the kink solution itself does not appear, as the operator is singular. 

More generally one expects such operators $\co$ to satisfy
\beq
[H,\co]|0\rangle=M\co|0\rangle \label{vuole}
\eeq
where $M$ is the mass.  When the theory is weakly coupled, $M$ should be equal to the mass of the corresponding classical solution plus quantum corrections.  

A general formalism for operators which create states whose form factors $\langle \phi(x)\rangle$ reproduce the classical solutions, and who solve (\ref{vuole}) was described in Ref.~\cite{taylor78}.    The general form of $\co$ is the product of a displacement operator, which fixes the form factor, with another operator which squeezes and deforms the state to minimize the energy in the displaced state.  This other operator can be calculated in perturbation theory. 

We would like to study these operators in weakly coupled theories, in which they correspond to classical solitons, with the hope that once we understand them in this context we may create them in strongly coupled theories in which they do not.  In the present paper we take a first step.  We will use the formalism of Ref.~\cite{taylor78} to rederive the leading correction to the quantum mass of the $\phi^4$ kink, first calculated in Ref.~\cite{dhn2}.  

Our method is particularly robust.  By normal-ordering our theory from the beginning, we render it finite\footnote{This is always the case with scalar field theories in 1+1 dimensions.}, so that divergences are never present in our calculation.   This eliminates the need to compactify and decompactify in the presence of an ultraviolet cutoff, as was done in \cite{dhn2}.  We find the mass by directly diagonalizing the Hamiltonian of the 1-kink sector.  This is done to subleading order in perturbation theory.  Our method produces not only all of the eigenvalues, but also all of the Hamiltonian eigenstates.

In Sec.~\ref{ptsez} we show that in the one kink sector, our Hamiltonian problem is equivalent to the P\"oschl-Teller problem.  Next in Sec.~\ref{solsez} we calculate the P\"oschl-Teller solutions, which are the eigenfunctions of our Hamiltonian.   These eigenfunctions are used to diagonalize the Hamiltonian in Sec.~\ref{diagsez}, yielding the kink mass.  Finally in Sec.~\ref{remsez} we describe the entire spectrum, the generalization to other classical solutions of other theories and also how these results may be used to construct the kink operator.

\begin{table}
\begin{center}
\begin{tabular}{|l|l|}
\hline
symbol&description\\
\hline
$H$&$\phi^4$ Hamiltonian \\
$H_0$&Free Hamiltonian \\
$H_{PT}$&P\"oschl-Teller Hamiltonian \\
$H\p$&Hamiltonian in the kink sector \\
$\tilde{T}_2$&P\"oschl-Teller potential term\\
$Q$&Leading quantum correction to kink mass \\
$E_K$&Energy of kink state \\
$f(x)$&Classical kink solution\\
$g(x)$&P\"oschl-Teller eigenfunctions\\
$\tilde{g}(p)$&Inverse Fourier transform of PT eigenfunctions\\
$C$&The normalization of $g$\\
$m$&$\sqrt{2\lambda}v$\\
$\beta$&$m/2$\\
$|\pm\rangle$&The two ground states\\
$|K\rangle$&The kink state\\
$\df$&Displacement operator, creating the form factor $f(x)$\\
$\co$&Operator $\co=\df\co_1$ that creates the kink state from $|-\rangle$\\
$a_p$&Annihilation operator for plane waves\\
$b_k$&Annihilation operator for continuous PT eigenstates\\
$b_{BO}$&Annihilation operator for odd bound PT eigenstate\\
$b_{BE}$&Annihilation operator for even bound PT eigenstate\\
\hline

\end{tabular}
\caption{Summary of Notation}
\end{center}
\end{table}

\section{The Modified P\"oschl-Teller Potential} \label{ptsez}

\subsection{A Ground State}

We begin with a real scalar field $\phi$ in 1+1 dimensions described by the Hamiltonian
\beq
H=\int dx \ch(x) \hsp
\ch(x)=\frac{1}{2}:\pi(x)\pi(x):+\frac{1}{2}:\partial_x\phi(x)\partial_x\phi(x):+\frac{\lambda}{4}:\left(\phi(x)-v\right)^2\left(\phi(x)+v\right)^2:
\eeq
where $\pi(x)$ is the conjugate momentum to $\phi(x)$ and $v$ and $\lambda$ are positive, real numbers.  As the field $\phi$ is tachyonic when expanded about zero, we will postpone our prescription for the normal ordering.  Note that the theory has two degenerate ground states
\beq
|\pm\rangle
\eeq
which satisfy
\beq
\langle\pm|\phi(x)|\pm\rangle=\pm v. \label{pheq}
\eeq

Let us consider for concreteness the ground state $|-\rangle$.  We may zero the expectation value of $\phi$ in Eq.~(\ref{pheq}) with the field redefinition
\beq
\phi\rightarrow\tilde{\phi}=\phi+v.
\eeq
From now on we will only be interested in $\tilde{\phi}$ and so we will drop the tildes.  Therefore now
\beq
\langle -|\phi(x)|-\rangle=0. 
\eeq

 In terms of this new field, the Hamiltonian is
\bea
H&=&H_0+H_1\hsp
H_0=\int dx \ch_0(x)\hsp
H_1=\int dx \ch_1(x)\nonumber\\
\ch_0(x)&=&\frac{1}{2}:\pi(x)\pi(x):+\frac{1}{2}:\partial_x\phi(x)\partial_x\phi(x):+\lambda v^2:\phi^2(x):\nonumber\\
\ch_1(x)&=&-\lambda v:\phi^3(x):+\frac{\lambda}{4}:\phi^4(x):. \label{h}
\eea
We can see that the new field $\phi(x)$ has a mass\footnote{Our $m$ differs by a factor of $\sqrt{2}$ from that of Ref.~\cite{dhn2}, who instead defined it to be the tachyonic mass of the unshifted vacuum.} of
\beq
m=\sqrt{2\lambda} v.
\eeq

As we work in 1+1 dimensions, $\phi$ is dimensionless and so $v$ is dimensionless while $\lambda$ has dimensions of $m^2$.  Therefore our perturbative expansion will be in $1/v$.  

It will be convenient to rewrite the Hamiltonian density in terms of $m$ and $\lambda$
\bea
\ch_0(x)&=&\frac{1}{2}:\pi(x)\pi(x):+\frac{1}{2}:\partial_x\phi(x)\partial_x\phi(x):+\frac{m^2}{2}:\phi^2(x):\nonumber\\
\ch_1(x)&=&-\frac{\sqrt{\lambda}m}{\sqrt{2}}:\phi^3(x):+\frac{\lambda}{4}:\phi^4(x):. \label{ch}
\eea
Now unfortunately our perturbative parameter has disappeared from the problem.  However, as $1/v$ is equal to $\sqrt{2\lambda}/m$, our expansion is equivalent to an expansion in $\sqrt{\lambda}$ with $m$ held fixed.

Although $\phi$ is not a free field, in the Schrodinger picture we can Fourier transform it to define oscillator modes $a$ and $a^\dag$
\beq
\phi(x)=\pin{p}\frac{1}{\sqrt{2\omega_p}}\left(a^\dag_p+a_{-p}\right)e^{-ipx}\hsp
\pi(x)=i\pin{p}\frac{\sqrt{\omega_p}}{\sqrt{2}}\left(a^\dag_p-a_{-p}\right)e^{-ipx} \label{osc}
\eeq
where
\beq
\omega_p=\sqrt{m^2+p^2}.
\eeq
The canonical commutation relations
\beq
[\phi(x),\pi(y)]=i\delta(x-y) \label{phipi}
\eeq
then yield
\beq
[a_p,a^\dag_q]=2\pi\delta(p-q).
\eeq
Finally we can define our normal ordering prescription:  All $a^\dag$ are placed on the left of all $a$. 

The $|-\rangle$ state can be calculated in perturbation theory in $\lambda$ in terms of the ground state of the free theory $H_0$, which is annihilated by all $a_p$.  It satisfies
\beq
H|-\rangle=E_0|-\rangle
\eeq
where $E_0$ is of order $\lambda$.  The state $|-\rangle$ can be constructed from the free ground state by acting with an operator which is equal to the identity plus corrections of order $\sqrt{\lambda}$.

\subsection{The Kink}

A single, time-independent, kink at rest corresponds to another state $|K\rangle$ which is also an eigenstate of the Hamiltonian
\beq
H|K\rangle=E_K|K\rangle. \label{scheq}
\eeq
We will refer to this equation as a Schrodinger equation\footnote{We hope that this terminology does not cause confusion, as we are working in quantum field theory and not quantum mechanics.}, and the quantity
\beq
M_K=E_K-E_0
\eeq
as the kink mass, as it is the minimal energy cost of creating a kink.  

Following the general arguments of Ref.~\cite{taylor78}, the kink state can be constructed as a coherent state by acting on $|-\rangle$ with the operator $\co$.  What do we know about $\co$?

The classical equations of motion for the field $\phi$ are
\beq
\frac{\partial^2\phi_{cl}(x,t)}{\partial t^2}-\frac{\partial^2\phi_{cl}(x,t)}{\partial x^2}=\frac{m^2}{2}\left(\phi_{cl}(x,t)-\frac{m}{\sqrt{2\lambda}}\right)-\lambda\left(\phi_{cl}(x,t)-\frac{m}{\sqrt{2\lambda}}\right)^3.
\eeq
One solution is the time independent kink
\beq
\phi_{cl}(x,t)=f(x)\hsp
f(x)=\frac{m}{\sqrt{2\lambda}}\left(1+{\rm{tanh}}\left(\frac{mx}{2}\right)\right). \label{ksol}
\eeq

This classical solution corresponds to a state in the quantum theory with
\beq
\langle K|\phi(x)|K\rangle=f(x)  \label{ff}
\eeq
plus quantum corrections.  How does one obtain such a state?

\subsection{The Displacement Operator}

Still following Ref.~\cite{taylor78}, we may obtain the form factor (\ref{ff}) using the dispacement operator
\beq
\df={\rm{exp}}\left(-i\int dx f(x)\pi(x)\right). \label{df}
\eeq
In this subsection the function $f(x)$ will be arbitrary, not necessarily a solution of the equations of motion.

The commutator with $\phi(x)$ may be obtained from
\beq
\left[\int dx f(x) \pi(x),\phi(y)\right]=\int dx f(x) \left[\pi(x),\phi(y)\right]=-if(y).
\eeq
As the right hand side is a scalar, it commutes with everything and so one easily obtains
\beq
[\df,\phi(y)]=-f(y)\df.
\eeq

From here one solution to Eq.~(\ref{ff}) is apparent.  If one defines
\beq
|f\rangle=\df|-\rangle
\eeq
then
\bea
\langle f|\phi(x)|f\rangle&=&\langle -|\df^\dag \phi(x)\df|-\rangle
=\langle -|\df^\dag [\phi(x),\df]|-\rangle+\langle -|\df^\dag\df \phi(x)|-\rangle\nonumber\\
&=&\langle -|\df^\dag\df f(x) |-\rangle+\langle -|\phi(x)|-\rangle=f(x)\langle - |-\rangle+0=f(x)
\eea
where we have assumed the state $|-\rangle$ to be normalized to unity and we used the unitarity of $\df$.  

We have not shown that $\co=\df$ but merely that $\df$ would yield the correct form factor (\ref{ff}).  More generally
\beq
\co=\df \co_1
\eeq
where $\co_1$ is another operator whose effect is subdominant in $\sqrt{\lambda}$.

Any change in the normal ordering prescription will affect the kink mass, and so we need to study the action of $\df$ on the normal ordering carefully.   Let us define the Fourier transform of $f(x)$ by
\beq
\tf(p)=\int dx f(x) e^{-ipx}.
\eeq

The commutators of the exponentials in Eq.~(\ref{df}) are
\beq
\left[\int dx f(x) \pi(x),a_q\right]=i\pin{p} \sqrt{\frac{\omega_p}{2}}\tf(p)\left[a^\dag_p,a_q\right]=-i \sqrt{\frac{\omega_q}{2}}\tf(q).
\eeq
and
\beq
\left[\int dx f(x) \pi(x),a^\dag_q\right]=-i\pin{p} \sqrt{\frac{\omega_p}{2}}\tf(p)\left[a_{-p},a^\dag_q\right]=-i \sqrt{\frac{\omega_q}{2}}\tf(-q).
\eeq
Again the right hand side is a scalar in both cases, and the commutators with the full exponential (\ref{df}) are easily calculated
\beq
\left[\df,a_q\right]=- \sqrt{\frac{\omega_q}{2}}\tf(q) \df\hsp
\left[\df,a^\dag_q\right]=- \sqrt{\frac{\omega_q}{2}}\tf(-q) \df.
\eeq

We will need to move products of $a_q$ and $a^\dag_q$ past $\df$.  From the commutators one finds that this is done by shifting $a_q$ and $a^\dag_q$
\beq
a_q\df=\df\left(a_q+\sqrt{\frac{\omega_q}{2}}\tf(q)\right)\hsp
a^\dag_q\df=\df\left(a^\dag_q+\sqrt{\frac{\omega_q}{2}}\tf(-q)\right). \label{af}
\eeq
Note that any normal ordered product will remain normal ordered when pushed past $\df$, as the substitution of $a_q$ or $a^\dag_q$ by a scalar leaves all $a^\dag$ on the left.  For example
\beq
:F(a^\dag_q,a_r):\df=\df:F\left(a^\dag_q+\sqrt{\frac{\omega_q}{2}}\tf(-q),a_r+\sqrt{\frac{\omega_r}{2}}\tf(r)\right): \label{aid}
\eeq
where $F$ is any function of two variables.  Similarly, as $\pi$ and $\df$ commute,
\beq
:F\left(\pi(x),\phi(x)\right):\df=\df:F\left(\pi(x),\phi(x)+f(x)\right):. \label{fident}
\eeq
Recall that the normal ordering prescription is always that of the decomposition of $\phi$ with mass $m$ into $a^\dag$ and $a$ as in Eq.~(\ref{osc}).  The identity (\ref{fident}) was derived without assuming that $f$ satisfies the equations of motion.

\subsection{Shifting the Hamiltonian}

To solve Eq.~(\ref{scheq}) for $E_K$, we will need to apply the identities in Eqs.~(\ref{aid}) and (\ref{fident}) to the Hamiltonian given in Eqs.~(\ref{h}) and (\ref{ch}).  Let us push $H$ past $\df$ one piece at a time
\bea
H_0\df&=&\pin{p}\omega_p a^\dag_p a_p \df=\df\pin{p}\omega_p \left(a^\dag_p+\sqrt{\frac{\omega_p}{2}}\tf(-p)\right)\left(a_p+\sqrt{\frac{\omega_p}{2}}\tf(p)\right)\nonumber\\
&=&\df \left(H_0 +\pin{p}\omega_p\sqrt{\frac{\omega_p}{2}}\tf(p)\left(a^\dag_p+a_{-p}\right)+\pin{p}\frac{\omega_p^2}{2}\tf(p)\tf(-p)\right)\nonumber\\
&=&\df \left(H_0+T_1+T_0\right) \label{h0f}
\eea
where $T_1$ and $T_0$ are the two terms in the previous expression.
Using
\beq
\omega_p^2=p^2+m^2\hsp
\omega_p^2\tf(p)=\int dx e^{-ixp}\left(m^2-\partial_x^2\right)f(x) \label{ftid}
\eeq
and
\beq
a^\dag_p+a_{-p}=\sqrt{2\omega_p}\int dy \phi(y) e^{ipy}
\eeq
we can simplify the second term in Eq.~(\ref{h0f})
\beq
T_1=\int dx \phi(x)\left(m^2-\partial_x^2\right)f(x).
\eeq
Similarly (\ref{ftid}) simplifies the third term to
\beq
T_0=\int dxf(x)\frac{m^2-\partial_x^2}{2}f(x). 
\eeq

Next we need to treat the interaction terms
\bea
H_1\df&=&\int dx \left(-\frac{\sqrt{\lambda}m}{\sqrt{2}}:\phi^3(x):+\frac{\lambda}{4}:\phi^4(x):\right)\df\nonumber\\
&=&\df\int dx \left(-\frac{\sqrt{\lambda}m}{\sqrt{2}}:(\phi(x)+f(x))^3:+\frac{\lambda}{4}:(\phi(x)+f(x))^4:\right)\nonumber\\
&=&\df \left(H_1+T\p_0+T\p_1+T\p_2+T\p_3\right)
\eea
where $T\p_N$ is of order $\phi^{N}$.  These are each easily evaluated
\beq
T\p_N=\int dx:\phi^N(x):\left(-\frac{3!}{N!(3-N)!}\frac{\sqrt{\lambda}m}{\sqrt{2}} f^{3-N}+\frac{4!}{N!(4-N)!}\frac{\lambda}{4}f^{4-N}\right) \label{tp}
\eeq
where it is understood that only terms with strictly positive powers of $f$ are included.

Putting $H_0$ and $H_1$ together, we are now ready to move the entire Hamiltonian past $\df$
\beq
H\df=\df \left(H+\tilde{T}_0+\tilde{T}_1+\tilde{T}_2+\tilde{T}_3\right)
\eeq
where
\beq
\tilde{T}_N=T\p_N+T_N
\eeq
is of order $\phi^N$.  Eq.~(\ref{tp}) yields the new interaction term $\tilde{T}_3$
\beq
\tilde{T}_3=T\p_3=\lambda\int dx f(x) :\phi^3(x):.
\eeq
The scalar term is
\beq
\tilde{T}_0=\int dx\left[\frac{m^2}{2} f^2(x)-\frac{f(x)f^{\prime\prime}(x)}{2}-\sqrt{\frac{\lambda}{2}}mf^3(x)+\frac{\lambda}{4}f^4(x)\right].
\eeq
Using the kink solution (\ref{ksol}) one finds
\beq
\tilde{T}_0=\frac{m^3}{3\lambda}
\eeq
which is the well-known formula for the classical energy of the kink.

The term linear in $\phi$ is
\beq
\tilde{T}_1=\int dx\phi(x)\left[m^2 f(x)-f^{\prime\prime}(x)-3\sqrt{\frac{\lambda}{2}}mf^2(x)+\lambda f^3(x)\right].
\eeq
As usual, the fact that $f(x)$ satisfies the classical equations of motion
\beq
f^{\prime\prime}(x)=-\frac{m^2}{2}\left(f(x)-\frac{m}{\sqrt{2\lambda}}\right)+\lambda\left(f(x)-\frac{m}{\sqrt{2\lambda}}\right)^3
\eeq
implies that the linear term vanishes
\beq
\tilde{T}_1=0.
\eeq
The most interesting term is the quadratic term
\beq
\tilde{T}_2=\int dx:\phi^2(x):\left[-3\sqrt{\frac{\lambda}{2}}mf(x)+\frac{3\lambda}{2} f^2(x)\right].
\eeq
Again using the solution (\ref{ksol}) one finds
\beq
\tilde{T}_2=-\frac{3m^2}{4}\int dx\ {\rm{sech}}^2\left(\frac{mx}{2}\right) :\phi^2(x):. \label{t2}
\eeq

Finally we may assemble our result
\beq
H\df=\df H\p\hsp
H\p=E_{cl}+H_{PT}+H_I \label{hdf}
\eeq
where the classical energy is
\beq
E_{cl}=\tilde{T}_0=\frac{m^3}{3\lambda} \label{ecl}
\eeq
the interaction terms are
\beq
H_I=H_1+\tilde{T}_3=\int dx\left[\left(-\sqrt{\frac{\lambda}{2}}m+\lambda f(x)\right) :\phi^3(x):+\frac{\lambda}{4}:\phi^4(x):
\right]
\eeq
and the remaining terms are
\beq
H_{PT}=\int dx\left[ \frac{:\pi^2(x):}{2}+\frac{:\partial_x\phi(x)\partial_x\phi(x):}{2}+\left(\frac{m^2}{2}-3\left(\frac{m}{2}\right)^2{\rm{sech}}^2\left(\frac{mx}{2}\right)\right):\phi^2(x):
\right]. \label{hpt}
\eeq
The mass term is space-dependent.   However it takes the form of the exactly solvable modified P\"oschl-Teller (PT) potential.

\subsection{A New Problem}

We wish to solve the Schrodinger equation (\ref{scheq}) for the kink state
\beq
|K\rangle=\df\co_1|-\rangle.
\eeq
We can now reorganize this equation using (\ref{hdf})
\bea
H|K\rangle&=&H\df\co_1|-\rangle=\df H\p\co_1|-\rangle\nonumber\\
&=&E_K|K\rangle=\df E_K\co_1|-\rangle.
\eea
Identifying the last term on each line, and using the fact that $\df$ is invertible, one finds
\beq
H\p\co_1|-\rangle=E_K\co_1|-\rangle.
\eeq
Subtracting the scalar $E_{cl}$ from both coefficients this yields our new problem
\beq
\left(H_{PT}+H_I\right)\co_1|-\rangle=(E_K-E_{cl})\co_1|-\rangle. \label{npr}
\eeq
We have reduced the problem (\ref{scheq}) of finding the kink state to a new problem (\ref{npr}).

What is this new problem?  It is a Schrodinger equation for the state $\co_1|-\rangle.$  As $\df$ is gone, there is no kink.  As a result, if desired, one could compactify the theory on a circle with periodic boundary conditions.  As we are searching for the kink ground state, the goal is to solve the eigenvalue problem such that $E_K$ is minimized.  Of course the global minimum would be to include $\mathcal{D}_{-f}$ so as to remove the kink.  To remove such a spurious solution, one should specify that the boundary conditions of $\langle\phi(x)\rangle$ are to be kept fixed during this minimization, which would be automatic were the theory compactified.

The $H_I$ operator may be treated using an ordinary perturbative expansion in $\sqrt{\lambda}$.  This is straightforward and will be done in a future work.  In the present paper we will solve the truncated problem
\beq
H_{PT}\co_1|-\rangle=(E_K-E_{cl})\co_1|-\rangle \label{trunc}
\eeq
which is sufficient to give the $O(m)$ contributions to $E_K$.    Therefore it will allow us to calculate $E_K$ to the same order as Ref.~\cite{dhn2}.   In this truncated problem the $\mathcal{D}_{-f}$ solution is not present.

Note that (\ref{trunc}) cannot be treated simply by perturbing about $H_0$ and expanding in powers of $\sqrt{\lambda}$, because there will be contributions with arbitrary numbers of mass terms $\tilde{T}_2$ which all contribute at the same order.  We have attempted this, and found that convergence is at best quite slow.  Instead, we will use the exact eigenfunctions of $H_{PT}$ to exactly diagonalize the Hamiltonian $H_{PT}$.

\section{Classical Solutions to the P\"oschl-Teller Problem} \label{solsez}

The PT theory is a free theory, in the sense that all terms in the Hamiltonian are at most quadratic in $\phi$.  However, due to the space-dependent mass term $\tilde{T}_2$, the solutions of the classical equations of motion are not plane waves\footnote{They are however plane waves asymptotically, as the mass term becomes constant far from the origin.}.  To solve the problem (\ref{trunc}) it will be convenient to decompose the field $\phi(x)$ into the basis of PT solutions with constant frequency.  In this way we will introduce creation and annihilation operators $b^\dag$ and $b$ which create and annihilate PT solutions.  The field $\phi(x)$ is the same quantum operator as it was in the previous section, the role of this new basis is simply to reorganize its projections so that the Hamiltonian is of the form $b^\dag b$.  This is useful because it implies that the ground state $\co_1|-\rangle$ in (\ref{trunc}) is the unique state annihilated by all operators $b$.   As this condition already completely characterizes the state, we will not need to find an explicit example of the operator $\co_1$. 

In this section we will calculate the inverse Fourier transforms of the eigenfunctions of the P\"oschl-Teller wave equation.  The reader who is not interested in this derivation may simply note that the answers are given in Eq.~(\ref{gtk}) for the continuum states and Eqs.~(\ref{gtbe}) and (\ref{gtbo}) for the even and odd bound states respectively.

\subsection{General Solutions}

The classical equation of motion derived from $H_{PT}$ in (\ref{hpt}) is
\beq
\partial^2_t\phi_{cl}(x,t)-\partial^2_x\phi_{cl}(x,t)=\left(-4\beta^2+6\beta^2{\rm{sech}}^2(\beta x)\right)\phi_{cl}(x,t)
\eeq
where for convenience we have defined
\beq
\beta=\frac{m}{2}.
\eeq
As we are looking for eigenstates of $H_{PT}$, the time-dependence should be of the form $e^{-i\omega t}$ and so we search for solutions of the form
\beq
\phi_{cl}(x,t)=f_k(x) e^{-i \omega_k t}.
\eeq
The functions $f_k$ then satisfy the equation
\beq
0=\partial^2_x f_k(x)+\left(\omega_k^2-4\beta^2+6\beta^2{\rm{sech}}^2(\beta x)\right)f_k(x). \label{fkeq}
\eeq
This can be recognized as the wave equation for a field in a well of width $1/\beta$.  The $6$ in (\ref{fkeq}) characterizes the depth of the potential well, and corresponds to the third reflectionless PT potential.  We will see how the fact that the potential is reflectionless affects the structure of the leading quantum correction to its mass.  The fact that it is the third means that there will be precisely two bound states.  

The term $\omega_k^2-4\beta^2$ is just energy squared minus mass squared, and so one would like it to be
\beq
\omega_k^2-4\beta^2=k^2. \label{kdef}
\eeq
As so far we have not defined our parametrization $k$, we will fix it by demanding (\ref{kdef}).  In general there will be two solutions with each value of $k^2$, one even and one odd.

\subsection{Continuum States}

After a change of variables
\beq
y=\cosh^2(\beta x)
\eeq
and dividing $f_k$ by $y^{3/2}$ our wave equation (\ref{fkeq}) becomes the hypergeometric equation and it has even and odd solutions \cite{flugge}
\bea
\psi^e_k(x)&=&\cosh^{3/2}(\beta x) F\left(\frac{3+ik/\beta}{2},\frac{3-ik/\beta}{2};\frac{1}{2};-\sinh^2(\beta x)\right) \label{gensol}\\
\psi^o_k(x)&=&\cosh^{3/2}(\beta x)\sinh(\beta x) F\left(\frac{4+ik/\beta}{2},\frac{4-ik/\beta}{2};\frac{3}{2};-\sinh^2(\beta x)\right)\nonumber
\eea
where $F$ are ordinary hypergeometric functions ${}_2F_1$ which are calculated in the Appendix.

Substituting (\ref{hyp}) into (\ref{gensol}) we find the solutions
\bea
\psi^e_k(x)&=&\left(1-\frac{3}{k^2/\beta^2+1}{\rm{\tanh}}^2(\beta x)\right)\cos(kx)-\frac{3k/\beta}{k^2/\beta^2+1}{\rm{tanh}}(\beta x)\sin(kx)\label{psi}\\
\psi^o_k(x)&=&\left(\frac{k^2/\beta^2+1-3{\rm{\tanh}}^2(\beta x)}{(k^2/\beta^2+4)k/\beta}\right)\sin(kx)+\frac{3}{k^2/\beta^2+4}{\rm{tanh}}(\beta x)\cos(kx).\nonumber
\eea
The function $\psi^e$ agrees with Ref.~\cite{lekner} while $\psi^o$ differs by a factor of $(k^2/\beta^2+4)$ in the first term.  We have checked that our functions satisfy the wave equation (\ref{fkeq}) and so we believe that our result is correct.  When $\beta|x|>>1$ the coefficients of $\sin(kx)$ and $\cos(kx)$ in Eq.~(\ref{psi}) are constant, and so the solutions are plane waves with wave number $k$, as expected far from the sech${}^2$ potential well.

The even and odd functions have different normalizations.  This can be fixed with a simple rescaling
\beq
\psi^e_k(x)\longrightarrow(k^2/\beta^2+1)\psi^e_k\hsp
\psi^o_k(x)\longrightarrow(k^2/\beta^2+4)k/\beta\psi^e_k
\eeq
which yields
\bea
\psi^e_k(x)&=&\left(k^2/\beta^2-2+3{\rm{sech}}^2(\beta x)\right)\cos(kx)-3k/\beta\ {\rm{tanh}}(\beta x)\sin(kx)\label{psi2}\\
\psi^o_k(x)&=&\left(k^2/\beta^2-2+3{\rm{sech}}^2(\beta x)\right)\sin(kx)+3k/\beta\ {\rm{tanh}}(\beta x)\cos(kx).\nonumber
\eea
The normalizations are now identical.  As these are eigenstates of a Hermitian Hamiltonian with distinct eigenvalues, the $\psi_k(x)$ at distinct $k$ are orthogonal.  The normalization can be obtained from the $\beta|x|>>1$ region, where all coefficients are constant
\beq
\int dx \psi^i_{k_1} (x) \psi^j_{k_2}(x)=\pi \delta^{ij} C^2_{k_1}\delta(k_1-k_2)\hsp
C_k=\sqrt{(k^2/\beta^2+1)(k^2/\beta^2+4)}\hsp i,j\in\{e,o\}. \label{normpsi}
\eeq
In the case of plane waves, the normalization constant analogous to $C_k$ was equal to unity.

We will need the inverse Fourier transforms of the wave functions.  As our answer differs from that obtained using Mathematica by some Dirac delta functions, we will derive our answer systematically here as we believe it to be correct.  Let us begin by decomposing $\psi^e_k$ into three pieces
\bea
\psi^e_k(x)&=&A^e_k(x)+B^e_k(x)+C^e_k(x)\hsp A^e_k(x)=\left(k^2/\beta^2-2\right)\cos(kx)\nonumber\\
B^e_k(x)&=&3\sech^2(\beta x)\cos(kx)\hsp
C^e_k(x)=-3k/\beta\ \tanh(\beta x)\sin(kx).
\eea

Contour integration, using Cauchy's theorem with residues evenly spaced along the imaginary axis, yields the inverse Fourier transform
\beq
\int dx \tanh(\beta x) e^{ipx}=\frac{\pi i}{\beta}\csch\left(\frac{\pi p}{2\beta}\right) \label{tanh}
\eeq
whose derivative is
\beq
\int dx \sech^2(\beta x)e^{ipx}=\frac{\pi p}{\beta^2}\csch\left(\frac{\pi p}{2\beta}\right). \label{sech}
\eeq
We will also need the identities
\bea
\int dx f(x) \cos(kx) e^{ipx}&=&\frac{1}{2} \int dx f(x) \left(e^{i(p+k)x}+e^{i(p-k)x}\right)=\frac{\tilde{f}(p+k)+\tilde{f}(p-k)}{2}\label{spost}\\
\int dx f(x) \sin(kx) e^{ipx}&=&\frac{1}{2i} \int dx f(x) \left(e^{i(p+k)x}-e^{i(p-k)x}\right)=\frac{\tilde{f}(p+k)-\tilde{f}(p-k)}{2i}\nonumber
\eea
where we have defined the inverse Fourier transform $\tilde{f}$ of an arbitrary function $f$ to be
\beq
\tilde{f}(p)=\int dx f(x) e^{ipx}.
\eeq

Combining these identities one finds the desired inverse Fourier transforms.  First
\beq
\tilde{A}^e_k(p)=(k^2/\beta^2-2)\int dx \cos(kx)e^{ipx}=(k^2/\beta^2-2)\pi\left(\delta(p+k)+\delta(p-k)\right)
\eeq
captures the asymptotic behavior of the eigenfunctions, which are just the same plane waves that one would find in a free theory.  Next using (\ref{sech}) and (\ref{spost})
\bea
\tilde{B}^e_k(p)&=&3\int dx \sech^2(\beta x)\cos(kx) e^{ipx}\nonumber\\
&=&\frac{3\pi}{2\beta^2}\left[(p+k)\csch\left(\frac{\pi (p+k)}{2\beta}\right)+(p-k)\csch\left(\frac{\pi (p-k)}{2\beta}\right)\right].
\eea
Finally combining  (\ref{tanh}) and (\ref{spost})
\bea
\tilde{C}^e_k(p)&=&-3k/\beta\int dx \tanh(\beta x)\sin(kx) e^{ipx}\nonumber\\
&=&\frac{3\pi k}{2\beta^2}\left[\csch\left(\frac{\pi (p-k)}{2\beta}\right)-\csch\left(\frac{\pi (p+k)}{2\beta}\right)\right]. \label{math}
\eea
In contrast Mathematica finds additional Dirac delta functions in (\ref{math}), as its calculation of Fourier transforms does not appear to respect (\ref{spost}) which follows from the shift invariance of the integral.  The same steps applied to $\psi^o_k$ yield the Fourier transforms
\bea
\tilde{A}^o_k(p)&=&(k^2/\beta^2-2)\pi i\left(\delta(p-k)-\delta(p+k)\right)\\
\tilde{B}^o_k(p)&=&\frac{3\pi}{2\beta^2} i \left[(p-k)\csch\left(\frac{\pi (p-k)}{2\beta}\right)-(p+k)\csch\left(\frac{\pi (p+k)}{2\beta}\right)\right]\nonumber\\
\tilde{C}^o_k(p)&=&\frac{3\pi k}{2\beta^2} i \left[\csch\left(\frac{\pi (p-k)}{2\beta}\right)+\csch\left(\frac{\pi (p+k)}{2\beta}\right)\right].\nonumber
\eea

We would like to simultaneously diagonalize $H_{PT}$ and the momentum $k$, and so we will assemble these even and odd real solutions into complex solutions
\beq
g_k(x)=\psi^e_k(x)-i\psi^o_k(x).
\eeq
The same decomposition into $A$, $B$ and $C$ may be applied to $g$ and its inverse Fourier transform, which by linearity of the inverse Fourier transform yields
\bea
\tilde{A}_k(p)&=&\tilde{A}^e_k(p)-i\tilde{A}^o_k(p)=(k^2/\beta^2-2)2\pi\delta(p-k)\\
\tilde{B}_k(p)&=&\frac{3\pi}{\beta^2} (p-k)\csch\left(\frac{\pi (p-k)}{2\beta}\right)\nonumber\\
\tilde{C}_k(p)&=&\frac{3\pi k}{\beta^2} \csch\left(\frac{\pi (p-k)}{2\beta}\right).\nonumber
\eea
Summing these we find our final answer for the inverse Fourier transform of the wave functions
\beq
\tilde{g}_k(p)=\tilde{A}_k(p)+\tilde{B}_k(p)+\tilde{C}_k(p)=(k^2/\beta^2-2)2\pi\delta(p-k)+\frac{3\pi p}{\beta^2} \csch\left(\frac{\pi (p-k)}{2\beta}\right). \label{gtk}
\eeq
The first term is the plane wave piece which comes from the fact that the eigenfunctions of $H_{PT}$ and $H_0$ are identical asymptotically.  The fact that there is no $\delta(p+k)$ term results from the reflectionless nature of the P\"oschl-Teller potential with coefficient equal to $n(n-1)/2$ for $n$ an integer.  The quantum corrections to the mass of the kink come from the second term, which has a simple pole at $p=k$ but is nonzero away from the pole.

Using the normalization (\ref{normpsi}) one easily finds
\beq
\int dx g^*_{k_1} (x) g_{k_2}(x)=2\pi C^2_{k_1}\delta(k_1-k_2) \label{norm}
\eeq
where the real and imaginary parts of $g_k(x)$ contribute equally.  We will also need the fact that
\beq
g^*_k(x)=g_k(-x)=g_{-k}(x).
\eeq
As a result of the first equality, $\tilde{g}_k(p)$ is real.  Also the inverse Fourier transforms satisfy
\beq
\tilde{g}_k(p)=\tilde{g}_{-k}(-p) \label{menogkp}
\eeq
and
\beq
\pin{p} {\tilde{g}}_{k_1} (p) {\tilde{g}}_{k_2}(p)=\int dx g_{k_1} (x) g_{k_2}(-x)=2\pi C^2_{k_1}\delta(k_1-k_2). \label{normp}
\eeq

\subsection{Bound States}

As the Hamiltonian is reflection-invariant, the nondegenerate bound states will be even or odd.  They are still given by (\ref{psi2}), however now the kinetic energy $\omega_k^2-4\beta^2$ is negative and so $k$ is imaginary.  Imposing that the wave function is normalizable at $|x|\rightarrow\infty$ yields only two bound states.  There is one even bound state $g_{BE}$ with $\omega=0$ and one odd bound state $g_{BO}$ with $\omega=\beta\sqrt{3}$.  This is well-known \cite{dhn2,flugge}.

Inserting
\beq
\omega_{BE}=0\hsp k_{BE}=2 i \beta 
\eeq
into the general solution (\ref{psi2}) for the even mode $\psi^e_k$ and dividing by $-3$ one finds the wave function of the even bound state
\beq
g_{BE}(x)=\sech^2(\beta x).
\eeq
This is proportional to the derivative of the classical kink solution (\ref{ksol}) and so we can identify it with the expected Goldstone mode corresponding to the translation symmetry broken by the kink.  The normalization is
\beq
\int dx g^2_{BE}(x)=C_{BE}^2\hsp C_{BE}=\frac{2}{\sqrt{3\beta}}.
\eeq
The inverse Fourier transform is given by Eq.~(\ref{sech})
\beq
\tilde{g}_{BE}(p)=\int dx g_{BE}(x) e^{ipx}=\frac{\pi p}{\beta^2}\csch\left(\frac{\pi p}{2\beta}\right).  \label{gtbe}
\eeq

Similarly, inserting 
\beq
\omega_{BO}=\beta\sqrt{3} \hsp k_{BO}=i \beta \label{wbo}
\eeq
into the solution (\ref{psi2}) for the odd mode $\psi^o_k$ and rescaling one finds the odd bound state
\beq
g_{BO}(x)=-i\frac{\sinh(\beta x)}{\cosh^2(\beta x)} \label{gbod}
\eeq
whose normalization is
\beq
\int dx g_{BO}(x)g^*_{BO}(x)=C_{BO}^2\hsp C_{BO}=\sqrt{\frac{2}{3\beta}}. \label{gbo}
\eeq
We included a factor of $-i$ in Eq.~(\ref{gbod}) because $g_{BO}(x)$ is odd, and our complex eigenfunctions are constructed from real even parts and imaginary odd parts so that their Fourier transforms will be real.  The inverse Fourier transform of $\sech$ can be evaluated by a contour integral whose residues are identical to those appearing in the transform of $\tanh$ up to relative signs and an overall phase, leading to the identity
\beq
\int dx \sech(\beta x) e^{ipx}=\frac{\pi}{\beta}\sech\left(\frac{\pi p}{2\beta}\right). \label{sech1}
\eeq
As Eq.~(\ref{gbo}) is proportional to the derivative of $\sech(\beta x)$, one finds
\beq
\tilde{g}_{BO}(p)=\int dx g_{BO}(x) e^{ipx}=\frac{\pi p}{\beta^2}  \sech\left(\frac{\pi p}{2\beta}\right). \label{gtbo}
\eeq
Note that
\beq
\tilde{g}_{BE}(-p)=\tilde{g}_{BE}(p)\hsp
\tilde{g}_{BO}(-p)=-\tilde{g}_{BO}(p). \label{dispari}
\eeq

\section{Mode Expansion of the P\"oschl-Teller Hamiltonian} \label{diagsez}

\subsection{PT Annihilation and Creation Operators}

This paper is about the dynamics of a quantum field $\phi(x)$.  The original Hamiltonian was the $\phi^4$ theory, but we found that the problem of finding the mass of the quantum kink is equivalent to another problem involving the PT Hamiltonian $H_{PT}$ plus interaction terms, which we have dropped as they are subdominant in our $\lambda$ expansion.  The Hamiltonian $H_{PT}$ is not equal to our original Hamiltonian $H$, but the quantum field is the same operator.

We know that $\phi(x)$ and its conjugate momentum can be expanded in oscillator modes $a^\dag_k$ and $a_k$, with an expansion given in (\ref{osc}).  This is an expansion in plane waves.  Our Hamiltonian $H_{PT}$ in Eq.~(\ref{hpt}) is the sum of two pieces
\beq
H_{PT}=H_0+\tilde{T}_2
\eeq
where $H_0$ is defined in Eqs.~(\ref{h}) and (\ref{ch}) and $\tilde{T}_2$ is defined in Eq.~(\ref{t2}).  While $H_0$ can be written in the form $a^\dag a$
\beq
H_0=\pin{p}\omega_p a^\dag_p a_p
\eeq
the same is not true of $\tilde{T}_2$, which contains terms $a^\dag_p a^\dag_{-p}$ and $a_pa_{-p}$.  As a result the ground state is probably not annihilated by all of the annihilation operators $a_p$, and so the Schrodinger equation (\ref{trunc}) is difficult to solve.  The problem of course is that the expansion (\ref{osc}) is an expansion in plane waves, which are eigenfunctions of $H_0$ but not of $H_{PT}$.  It is in the basis of the latter that the Hamiltonian is diagonal and so the ground state corresponds to zero excitations, and so is annihilated by the corresponding annihilation operators.

This motivates us to instead expand $\phi(x)$ in terms of the eigenfunctions of $H_{PT}$.  We have seen that there are three kinds of eigenfunctions: continuum eigenfunctions and an odd and even bound state.  Thus we will decompose $\phi(x)$ and $\pi(x)$ into three pieces
\beq
\phi(x)=\phi_C(x)+\phi_{BO}(x)+\phi_{BE}(x)\hsp
\pi(x)=\pi_C(x)+\pi_{BO}(x)+\pi_{BE}(x).
\eeq
Now we will define our PT annihilation and creation operators similarly to the free case, but using the eigenstates $g(x)$ of $H_{PT}$ instead of the plane wave eigenstates of $H_0$
\bea
\phi_C(x)&=&\pin{k}\frac{1}{\sqrt{2\omega_k}}\left(b_k^\dag+b_{-k}\right)\frac{g_k(x)}{C_k}\nonumber\\
\phi_{BO}(x)&=&\frac{1}{\sqrt{2\omega_{BO}}}\left(b_{BO}^\dag-b_{BO}\right)\frac{g_{BO}(x)}{C_{BO}}\nonumber\\
\phi_{BE}(x)&=&\phi_0 \frac{g_{BE}(x)}{C_{BE}}. \label{phib}
\eea
Note that $\phi_{BE}$, corresponding to the Goldstone mode, could not be defined similarly to the others because $\omega_{BE}=0$.  Therefore we have defined a new operator $\phi_0$ instead of introducing oscillators $b$ and $b^\dag$.  Also we have chosen a relative minus sign in our definition of $b_{BO}$.  This is is necessary to arrive at the canonical commutation relations for $b_{BO}$.  Intuitively it is necessary because $g_{BO}(x)$ is odd in $k$, and so this is the natural generalization of the sign choices in the definition of $b_k$.

We similarly decompose the conjugate momentum field
\bea
\pi_C(x)&=&i \pin{k}\sqrt{\frac{\omega_k}{2}}\left(b_k^\dag - b_{-k}\right)\frac{g_k(x)}{C_k}\nonumber\\
\pi_{BO}(x)&=&i \sqrt{\frac{\omega_{BO}}{2}}\left(b_{BO}^\dag+b_{BO}\right)\frac{g_{BO}(x)}{C_{BO}}\nonumber\\
\pi_{BE}(x)&=&\pi_0 \frac{g_{BE}(x)}{C_{BE}} \label{pib}
\eea
where we have introduced the operator $\pi_0$ for the momentum of the Goldstone mode.  Note that Eqs.~(\ref{phib}) and (\ref{pib}) are merely definitions of the fields $b$, $b^\dag$, $\phi_0$ and $\pi_0$ as expansions of the field $\phi(x)$ and its conjugate $\pi(x)$ in the basis given by the $g(x)$.  We have not yet used the Hamiltonian or the fact that the $g$ are eigenstates.

Using the completeness of the eigenfunctions $g(x)$,  these relations can be inverted to provide explicit definitions of our new operators.  For the continuum states
\beq
b^\dag_k=\int dx \left[ \sqrt{\frac{\omega_k}{2}}\phi(x)-\frac{i}{\sqrt{2\omega_k}}\pi(x)\right]\frac{g^*_k(x)}{C_k}\hsp
b_{-k}=\int dx \left[ \sqrt{\frac{\omega_k}{2}}\phi(x)+\frac{i}{\sqrt{2\omega_k}}\pi(x)\right]\frac{g^*_k(x)}{C_k}
\eeq
from which the canonical commutation relations (\ref{phipi}) of $\phi(x)$ and $\pi(x)$ together with the normalization (\ref{norm}) yield the commutation relations of the new oscillators
\beq
[b_{k_1},b^\dag_{k_2}]=2\pi\delta(k_1-k_2).
\eeq
Similarly for the odd bound state
\bea
b_{BO}^\dag&=&\int dx \left[ \sqrt{\frac{\omega_{BO}}{2}}\phi(x)-\frac{i}{\sqrt{2\omega_{BO}}}\pi(x)\right]\frac{g^*_{BO}(x)}{C_{BO}}\nonumber\\
b_{BO}&=&\int dx \left[ -\sqrt{\frac{\omega_{BO}}{2}}\phi(x)-\frac{i}{\sqrt{2\omega_{BO}}}\pi(x)\right]\frac{g^*_{BO}(x)}{C_{BO}}
\eea
which, using the fact that $g_{BO}$ is imaginary, yields
\beq
[b_{BO},b^\dag_{BO}]=1.
\eeq
Finally, for the even bound state,
\beq
\phi_0=\int dx \phi(x)\frac{g^*_{BE}(x)}{C_{BE}}\hsp
\pi_0=\int dx \pi(x)\frac{g^*_{BE}(x)}{C_{BE}}. \label{pi0int}
\eeq
The complex conjugation is not important here as $g_{BE}(x)$ is real.
From (\ref{pi0int}) we see that the field and momentum zero modes satisfy the canonical commutations
\beq
[\phi_0,\pi_0]=i.
\eeq
Therefore $\phi_0$ and $\pi_0$, unlike the $b$ operators, do not create and annihilate excitations.  Rather they are the position and momentum operators for the kink.

We would like to rewrite the Hamiltonian in terms of the new oscillators $b$, $b^\dag$, $\phi_0$ and $\pi_0$.  However we cannot simply substitute Eqs.~(\ref{phib}) and (\ref{pib}) into our formula (\ref{hpt}) for the Hamiltonian because the latter is normal ordered in terms of $a^\dag$ and $a$.  The new oscillator modes will not be normal ordered, and in fact it is precisely this failure of normal ordering which is responsible for the quantum mass of the kink.

Therefore we will proceed as follows.  First we will write the Hamiltonian in terms of $a$ and $a^\dag$, where the normal-ordering is easily achieved.  Then we will rewrite $a$ and $a^\dag$ in terms of the PT oscillator modes.  To do this, we note that Eq.~(\ref{osc}) is easily inverted to obtain
\beq
a^\dag_p=\int dx \left[ \sqrt{\frac{\omega_p}{2}}\phi(x)-\frac{i}{\sqrt{2\omega_p}}\pi(x)\right]e^{ipx}\hsp
a_{-p}=\int dx \left[ \sqrt{\frac{\omega_p}{2}}\phi(x)+\frac{i}{\sqrt{2\omega_p}}\pi(x)\right]e^{ipx}. \label{phia}
\eeq
We will decompose these into their projections onto PT eigenfunctions
\beq
a^\dag_p=a^\dag_{C,p}+a^\dag_{BO,p}+a^\dag_{BE,p}\hsp
a_p=a_{C,p}+a_{BO,p}+a_{BE,p}
\eeq
which are found by inserting (\ref{phib}) and (\ref{pib}) into Eq.~(\ref{phia})
\bea
a^\dag_{C,p}&=&\pin{k}\frac{\tilde{g}_k(p)}{2C_k}\left(\frac{\omega_p+\omega_k}{\sqrt{\omega_p\omega_k}}b_k^\dag+\frac{\omega_p-\omega_k}{\sqrt{\omega_p\omega_k}}b_{-k}\right) \label{bog}\\
a_{C,-p}&=&\pin{k}\frac{\tilde{g}_k(p)}{2C_k}\left(\frac{\omega_p-\omega_k}{\sqrt{\omega_p\omega_k}}b_k^\dag+\frac{\omega_p+\omega_k}{\sqrt{\omega_p\omega_k}}b_{-k}\right)\nonumber\\
a^\dag_{BO,p}&=&\frac{\tilde{g}_{BO}(p)}{2C_{BO}}\left(\frac{\omega_p+\omega_{BO}}{\sqrt{\omega_p\omega_{BO}}}b_{BO}^\dag-\frac{\omega_p-\omega_{BO}}{\sqrt{\omega_p\omega_{BO}}}b_{BO}\right)\nonumber\\
a_{BO,-p}&=&\frac{\tilde{g}_{BO}(p)}{2C_{BO}}\left(\frac{\omega_p-\omega_{BO}}{\sqrt{\omega_p\omega_{BO}}}b_{BO}^\dag-\frac{\omega_p+\omega_{BO}}{\sqrt{\omega_p\omega_{BO}}}b_{BO}\right)\nonumber\\
a^\dag_{BE,p}&=&\frac{\tilde{g}_{BE}(p)}{C_{BE}}\left[ \sqrt{\frac{\omega_p}{2}}\phi_0-\frac{i}{\sqrt{2\omega_p}}\pi_0\right]\hsp 
a_{BE,-p}=\frac{\tilde{g}_{BE}(p)}{C_{BE}}\left[ \sqrt{\frac{\omega_p}{2}}\phi_0+\frac{i}{\sqrt{2\omega_p}}\pi_0\right].\nonumber
\eea
These are essentially Bogoliubov transformations, although they would be of the standard form only were $\tilde{g}_k(p)$ supported on $p=\pm k$.

\subsection{Continuum State Contribution}

All that remains to do is insert (\ref{bog}) into our Hamiltonian $H_{PT}$ to rewrite it as a free theory whose Schrodinger equation we may trivially solve by turning off all $b$ oscillators as well as the kink momentum $\pi_0$.  Let us start by decomposing $H_0$ into parts with contributions from distinct PT eigenfunctions
\beq
H_0=H_{C,0}+H_{BO,0}+H_{BE,0}.
\eeq
In principle there may be cross terms, in which for example both $a^\dag_C$ and $a_{BO}$ appear.  However such cross-terms vanish due to the orthogonality of the eigenfunctions $g(x)$.

Now we can calculate the continuous contribution
\bea
H_{C,0}&=&\pin{p} \omega_p a^\dag_{C,p} a_{C,p}\\
&=&\frac{1}{4}\pin{p}\pin{k_1}\pin{k_2}\frac{\tilde{g}_{k_1}(p)\tilde{g}_{k_2}(-p)}{C_{k_1}C_{k_2}\sqrt{\omega_{k_1}\omega_{k_2}}}\nonumber\\
&\times&\left[(\omega_p^2-\omega_{k_1}\omega_{k_2})(b^\dag_{k_1}b^\dag_{k_2}+b_{-k_1}b_{-k_2})+2(\omega_p^2+\omega_{k_1}\omega_{k_2})b^\dag_{k_1}b_{-k_2}\right.\nonumber\\
&&\left.+(\omega_p-\omega_{k_1})(\omega_p-\omega_{k_2})[b_{-k_2},b^\dag_{k_1}]\right]\nonumber\\
&=&\frac{1}{4}\pin{k_1}\pin{k_2}\frac{1}{C_{k_1}C_{k_2}\sqrt{\omega_{k_1}\omega_{k_2}}}\nonumber\\
&\times&\left[(I_3(k_1,k_2)-I_4(k_1,k_2))(b^\dag_{k_1}b^\dag_{k_2}+b_{-k_1}b_{-k_2})+2(I_3(k_1,k_2)+I_4(k_1,k_2))b^\dag_{k_1}b_{-k_2}\right]\nonumber\\
&&+\frac{1}{4}\pin{k}\frac{I_5(k)}{C_k^2\omega_k}\nonumber
\eea
where we have used the $k_1\leftrightarrow k_2$ symmetry to simplify the first term.  We have defined the integrals over $p$
\bea
I_3(k_1,k_2)&=&\pin{p}\omega_p^2\tilde{g}_{k_1}(p)\tilde{g}_{k_2}(-p)\hsp
I_4(k_1,k_2)=\pin{p}\omega_{k_1}\omega_{k_2}\tilde{g}_{k_1}(p)\tilde{g}_{k_2}(-p)\nonumber\\
I_5(k)&=&\pin{p}(\omega_p-\omega_k)^2\tilde{g}_k(p)\tilde{g}_{k}(p)
\eea
where we have used (\ref{menogkp}) to remove two minus signs in $I_5$.

Using the normalization
\beq
\pin{p} \tilde{g}_{k_1}(p)\tilde{g}_{k_2}(-p)=2\pi C_{k_1}^2\delta(k_1+k_2)
\eeq
one easily evaluates $I_4$
\beq
I_4(k_1,k_2)=2\pi C_{k_1}^2\omega_{k_1}^2\delta(k_1+k_2).
\eeq
The integral $I_3$ may be simplified by Fourier transforming and using the equations of motion (\ref{fkeq}) which are satisfied by $g_k(x)$
\bea
I_3(k_1,k_2)&=&\pin{p}\int dx \int dy g_{k_1}(x)g_{k_2}(y)e^{ip(x-y)}(4\beta^2+p^2)\\
&=&\pin{p}\int dx \int dy g_{k_1}(x)g_{k_2}(y)(4\beta^2-\partial_y^2)e^{ip(x-y)}\nonumber\\
&=&\pin{p}\int dx \int dy g_{k_1}(x)e^{ip(x-y)}(4\beta^2-\partial_y^2)g_{k_2}(y)\nonumber\\
&=&\pin{p}\int dx \int dy g_{k_1}(x)e^{ip(x-y)}(\omega_{k_2}^2+6\sech^2(\beta y))g_{k_2}(y)\nonumber\\
&=&\int dx (\omega_{k_2}^2+6\beta^2\sech^2(\beta x)) g_{k_1}(x)g_{k_2}(x)\nonumber\\
&=&I_4(k_1,k_2)+6\beta^2\int dx\ \sech^2(\beta x) g_{k_1}(x)g_{k_2}(x).\nonumber
\eea

Assembling these contributions
\bea
H_{C,0}&=&\frac{1}{4}\pin{k}\frac{I_5(k)}{C_k^2\omega_k}+\frac{3\beta^2}{2}\int dx\pin{k_1}\pin{k_2}\sech^2(\beta x)\frac{g_{k_1}(x)g_{k_2}(x)}{C_{k_1}C_{k_2}\sqrt{\omega_{k_1}\omega_{k_2}}}(b^\dag_{k_1}b^\dag_{k_2}+b_{-k_1}b_{-k_2})\nonumber\\
&&+\pin{k}\omega_k b^\dag_k b_k+3\beta^2\int dx\pin{k_1}\pin{k_2}\sech^2(\beta x)\frac{g_{k_1}(x)g_{k_2}(x)}{C_{k_1}C_{k_2}\sqrt{\omega_{k_1}\omega_{k_2}}}b^\dag_{k_1} b_{-k_2}. \label{hco}
\eea
The first terms on each line are the kind that we expect.  The first term in the first line is a scalar, and so contributes to the vacuum energy of the model, which is our quantum kink mass.  The first term on the second line is the expected oscillator sum in a free theory.  

The other terms should not be present in $H_{PT}$ as it should also be a noninteracting theory.    However so far we have only calculated the continuous contribution to $H_0$.  We must also add the continuum contribution to the PT potential $\tilde{T}_2$.  We decompose it as was done for $H_0$
\beq
\tilde{T}_2=\tilde{T}_{C,2}+\tilde{T}_{BO,2}+\tilde{T}_{BE,2}.
\eeq
The continuum term is
\bea
\tilde{T}_{C,2}&=&-3\beta^2\int dx\ {\rm{sech}}^2\left(\beta x\right) :\phi^2_C(x):\\
&=&-\frac{3\beta^2}{2}\int dx\pin{p}\pin{q}\frac{\sech^2(\beta x)}{\sqrt{\omega_p\omega_q}}e^{-i(p+q)x}\left(a^\dag_{C,p} a^\dag_{C,q}+a^\dag_{C,p} a_{C,-q}+a^\dag_{C,q}a_{C,-p}+a_{C,-q}a_{C,-p}\right)\nonumber\\
&=&-\frac{3\beta^2}{8}\int dx \pin{p}\pin{q} \frac{\sech^2(\beta x)}{\omega_p\omega_q}e^{-i(p+q)x}\pin{k_1}\pin{k_2}\frac{\tilde{g}_{k_1}(p)\tilde{g}_{k_2}(q)}{C_{k_1}C_{k_2}\sqrt{\omega_{k_1}\omega_{k_2}}}\nonumber\\
&\times&\left[4\omega_p\omega_q(b^\dag_{k_1}b^\dag_{k_2}+b_{-k_1}b_{-k_2})+2\omega_q(2\omega_p+\omega_{k_1}+\omega_{k_2})b^\dag_{k_1}b_{-k_2}+2\omega_q(2\omega_p-\omega_{k_1}-\omega_{k_2})b_{-k_2}b^\dag_{k_1}\right]\nonumber\\
&=&A+B\nonumber
\eea
where $A$ contains all terms with $b^\dag b^\dag$ and $bb$ while $B$ contains the others.  Note that $A$ cancels precisely with the corresponding terms in Eq.~(\ref{hco}).  This means that $H_{C,0}$ is of the form $b^\dag b$ plus a constant.  This simplification is the reason that we introduced the $b$ oscillators.

Let us simplify $B$ by rewriting the $b^\dag b$ and $b b^\dag$ terms as $b^\dag b$ terms and commutator terms, which are scalars
\bea
B&=&-3\beta^2\int dx \pin{p}\pin{q} \sech^2(\beta x)e^{-i(p+q)x}\pin{k_1}\pin{k_2}\frac{\tilde{g}_{k_1}(p)\tilde{g}_{k_2}(q)}{C_{k_1}C_{k_2}\sqrt{\omega_{k_1}\omega_{k_2}}}b^\dag_{k_1}b_{-k_2}\nonumber\\
&&-\frac{3\beta^2}{2}\int dx \pin{p}\pin{q} \sech^2(\beta x)e^{-i(p+q)x}\pin{k_1}\pin{k_2}\frac{\tilde{g}_{k_1}(p)\tilde{g}_{k_2}(q)}{C_{k_1}C_{k_2}\sqrt{\omega_{k_1}\omega_{k_2}}}[b_{-k_2},b^\dag_{k_1}]\nonumber\\
&&+\frac{3\beta^2}{4}\int dx \pin{p}\pin{q} \frac{\sech^2(\beta x)}{\omega_p}e^{-i(p+q)x}\pin{k_1}\pin{k_2}\frac{\tilde{g}_{k_1}(p)\tilde{g}_{k_2}(q)}{C_{k_1}C_{k_2}\sqrt{\omega_{k_1}\omega_{k_2}}}\nonumber\\
&&\times(\omega_{k_1}+\omega_{k_2})[b_{-k_2},b^\dag_{k_1}]\nonumber\\
&=&-3\beta^2\int dx\ \sech^2(\beta x)\pin{k_1}\pin{k_2}\frac{{g}_{k_1}(x){g}_{k_2}(x)}{C_{k_1}C_{k_2}\sqrt{\omega_{k_1}\omega_{k_2}}}b^\dag_{k_1}b_{-k_2}\nonumber\\
&&-\frac{3\beta^2}{2}\int dx\ \sech^2(\beta x) \pin{k}\frac{{g}_{k}(x)g^*_k(x)}{C_{k}^2\omega_{k}}\nonumber\\
&&+\frac{3\beta^2}{2}\int dx\pin{p}\pin{q} \frac{\sech^2(\beta x)}{\omega_p}e^{-i(p+q)x}\pin{k}\frac{\tilde{g}_{k}(p)\tilde{g}_{-k}(q)}{C_{k}^2}.
\eea
The $b^\dag b$ term cancels that in (\ref{hco}), leaving the last two lines, which are scalars.

Summarizing, the continuum contribution to the Hamiltonian is
\beq
H_C=\pin{k}\omega_k b^\dag_k b_k+Q_C
\eeq
where the scalar term $Q_C$ is
\bea
Q_C&=&\frac{1}{4}\pin{k}\frac{I_5(k)}{C_k^2\omega_k}+\frac{3\beta^2}{2}\int dx\pin{p}\pin{q} \frac{\sech^2(\beta x)}{\omega_p}e^{-i(p+q)x}\pin{k}\frac{\tilde{g}_{k}(p)\tilde{g}_{-k}(q)}{C_{k}^2}\nonumber\\
&&-\frac{3\beta^2}{2}\int dx\ \sech^2(\beta x) \pin{k}\frac{{g}_{k}(x)g^*_k(x)}{C_{k}^2\omega_{k}}.
\eea
Let us rewrite $Q_C$ in a mixed position-momentum form
\bea
Q_C&=&\frac{1}{4}\pin{k}\frac{I_5(k)}{C_k^2\omega_k}+\frac{3\beta^2}{2}\int dx\pin{p} \frac{\sech^2(\beta x)}{\omega_p}e^{-ipx}\pin{k}\frac{{\tilde{g}}_{k}(p)g_{-k}(x)}{C_{k}^2}\nonumber\\
&&-\frac{3\beta^2}{2}\int dx\pin{p}\ \sech^2(\beta x) e^{-ipx}\pin{k}\frac{{g}_{-k}(x){\tilde{g}}_{k}(p)}{C_{k}^2\omega_{k}}.
\eea
Now the equations of motion imply
\bea
6\beta^2\int dx\pin{p}e^{-ipx}\sech^2(\beta x) g_{-k}(x) &=&\int dx\pin{p}e^{-ipx} (-k^2-\partial_x^2) g_{-k}(x)\label{eomid}\\
&=& \int dx\pin{p}g_{-k}(x)(-k^2-\partial_x^2)e^{-ipx}\nonumber\\
&=& \int dx\pin{p}g_{-k}(x)(p^2-k^2)e^{-ipx}\nonumber\\
&=&\int dx\pin{p}g_{-k}(x)(\omega_p^2-\omega_k^2)e^{-ipx}\nonumber\\
&=&\pin{p}{\tilde{g}}_{-k}(-p)(\omega_p^2-\omega_k^2).
\nonumber
\eea
This derivation also works with the integrand multiplied by any function of $p$ but not $x$.  
Eq.~(\ref{eomid}) allows us to remove the $\sech$ from $Q_C$
\bea
Q_C&=&\frac{1}{4}\pin{k}\pin{p}\left[\frac{(\omega_p-\omega_k)^2}{\omega_k}+\frac{\omega_p^2-\omega_k^2}{\omega_p}-\frac{\omega_p^2-\omega_k^2}{\omega_k}
\right]\frac{\tilde{g}^2_{k}(p)}{C_{k}^2}\nonumber\\
&=&-\frac{1}{4}\pin{k}\pin{p}\frac{(\omega_p-\omega_k)^2}{\omega_p}\frac{\tilde{g}^2_{k}(p)}{C_{k}^2} . \label{qc}
\eea
This is the main result of this subsection.  As $\tilde{g}$ is real, it is real.  Notice that the integrand is nonsingular because each $\tilde{g}_k(p)$ has a simple pole at $k=p$ and the total double pole is cancelled by the double zero in $(\omega_p-\omega_k)^2$.   Similarly the delta functions in $\tilde{g}$ only appear with zero coefficient, whereas a nonzero coefficient may have led to a divergence.  $Q_C$ is the contribution to the kink energy of the continuum $PT$ modes.

The fact that $Q_C$ is negative is a result of the fact that the PT potential is negative.  This is the first quantum correction to the energy resulting from the existence of a potential well, and so it must be negative.  The $\omega_p$ in the denominator never vanishes due to the mass gap.  On the contrary, had the $\omega_k$ in the denominator of $H_{C,0}$ not been cancelled by $\tilde{T}_2$, it would have been a problem later when we consider the Goldstone mode $H_{BE,0}$, which has $\omega_k=0$.

The best feature of this expression is that every trace of the potential $\sech$ has disappeared.  They have disappeared not because of any nice property of the $\sech$ function, but just because we have used the equations of motion to replace them with the momentum squared.  This leads us to believe that had we chosen any other classical solution in a 1+1 dimensional theory with a canonical kinetic term, we could have done exactly the same manipulations, replacing the new potential with the momentum squared, and so obtained the same answer in terms of the eigenfunctions for its potential.   Thus we conjecture that (\ref{qc}) applies to all time-independent classical solutions in such theories.

\subsection{Odd Bound State Contribution}

The formulas for the odd bound state are essentially the same as that for the continuum, but without the index $k$ and with an extra minus sign before every $b_{BO}$ but not $b^\dag_{BO}$.

Let us start with $H_{BO,0}$.  Now we can calculate the continuous contribution
\bea
H_{BO,0}&=&\pin{p} \omega_p a^\dag_{BO,p} a_{BO,p}\\
&=&\frac{1}{4}\pin{p}\frac{\tilde{g}_{BO}(p)\tilde{g}_{BO}(-p)}{C_{BO}^2\omega_{BO}}\nonumber\\
&\times&\left[(\omega_p^2-\omega_{BO}^2)(b^\dag_{BO}b^\dag_{BO}+b_{BO}b_{BO})-2(\omega_p^2+\omega_{BO}^2)b^\dag_{BO}b_{BO}-(\omega_p-\omega_{BO})^2[b_{BO},b^\dag_{BO}]\right]\nonumber\\
&=&\frac{1}{4C_{BO}^2\omega_{BO}}\left[(I_0-I_1)(b^\dag_{BO}b^\dag_{BO}+b_{BO}b_{BO})-2(I_0+I_1)b^\dag_{BO}b_{BO}\right]+\frac{1}{4}\frac{I_2}{C_{BO}^2\omega_{BO}}.\nonumber
\eea
We have defined the integrals over $p$
\bea
I_0&=&\pin{p}\omega_p^2\tilde{g}_{BO}(p)\tilde{g}_{BO}(-p)\hsp
I_1=\pin{p}\omega_{BO}^2\tilde{g}_{BO}(p)\tilde{g}_{BO}(-p)\nonumber\\
I_2&=&\pin{p}(\omega_p-\omega_{BO})^2\tilde{g}_{BO}(p)\tilde{g}_{BO}(p)
\eea
where we have used (\ref{dispari}) to remove two minus signs in $I_2$.

Using the normalization
\beq
\pin{p} \tilde{g}_{BO}(p)\tilde{g}_{BO}(-p)=-C_{BO}^2
\eeq
one finds 
\beq
I_1=-C_{BO}^2\omega_{BO}^2.
\eeq
The integral $I_0$ may be simplified as was done above for $I_3$
\bea
I_0&=&\pin{p}\int dx \int dy g_{BO}(x)g_{BO}(y)e^{ip(x-y)}(4\beta^2+p^2)\\
&=&\int dx (\omega_{BO}^2+6\beta^2\sech^2(\beta x)) g_{BO}(x)g_{BO}(x)\nonumber\\
&=&I_1+6\beta^2\int dx\ \sech^2(\beta x) g_{BO}(x)g_{BO}(x).\nonumber
\eea

Assembling these contributions
\bea
H_{BO,0}&=&\frac{1}{4}\pin{k}\frac{I_2}{C_{BO}^2\omega_{BO}}+\frac{3\beta^2}{2}\int dx\sech^2(\beta x)\frac{g_{BO}(x)g_{BO}(x)}{C_{BO}^2\omega_{BO}}(b^\dag_{BO}b^\dag_{BO}+b_{BO}b_{BO})\nonumber\\
&&+\omega_{BO} b^\dag_{BO} b_{BO}-3\beta^2\int dx\sech^2(\beta x)\frac{g_{BO}(x)g_{BO}(x)}{C_{BO}^2\omega_{BO}}b^\dag_{BO} b_{BO}. \label{hbo}
\eea
Again the first terms on the first two lines are scalar contributions to the kink mass and also the free theory oscillator term respectively, while we expect other terms to be canceled by the potential term $\tilde{T}_2$.

Next we evaluate the odd bound contribution to $\tilde{T}_2$
\bea
\tilde{T}_{BO,2}&=&-\frac{3\beta^2}{2}\int dx\pin{p}\pin{q}\frac{\sech^2(\beta x)}{\sqrt{\omega_p\omega_q}}e^{i(p+q)x}\\
&&\times \left(a^\dag_{BO,p} a^\dag_{BO,q}+a^\dag_{BO,p} a_{BO,-q}+a^\dag_{BO,q}a_{BO,-p}+a_{BO,-q}a_{BO,-p}\right)\nonumber\\
&=&-\frac{3\beta^2}{8}\int dx \pin{p}\pin{q} \frac{\sech^2(\beta x)}{\omega_p\omega_q}e^{i(p+q)x}\frac{\tilde{g}_{BO}(p)\tilde{g}_{BO}(q)}{C_{BO}^2\omega_{BO}}\nonumber\\
&\times&\left[4\omega_p\omega_q(b^\dag_{BO}b^\dag_{BO}+b_{BO}b_{BO})-2\omega_q(2\omega_p+2\omega_{BO})b^\dag_{BO}b_{BO}-2\omega_q(2\omega_p-2\omega_{BO})b_{BO}b^\dag_{BO}\right]\nonumber\\
&=&A+B\nonumber
\eea
where $A$ again contains all terms with $b^\dag b^\dag$  and $bb$ and precisely cancels with the corresponding terms in Eq.~(\ref{hbo}).  

Let us simplify $B$ as in the continuous case
\bea
B&=&3\beta^2\int dx \pin{p}\pin{q} \sech^2(\beta x)e^{i(p+q)x}\frac{\tilde{g}_{BO}(p)\tilde{g}_{BO}(q)}{C_{BO}^2\omega_{BO}}b^\dag_{BO}b_{BO}\nonumber\\
&&+\frac{3\beta^2}{2}\int dx \pin{p}\pin{q} \sech^2(\beta x)e^{i(p+q)x}\frac{\tilde{g}_{BO}(p)\tilde{g}_{BO}(q)}{C_{BO}^2\omega_{BO}}[b_{BO},b^\dag_{BO}]\nonumber\\
&&-\frac{3\beta^2}{4}\int dx \pin{p}\pin{q} \frac{\sech^2(\beta x)}{\omega_p}e^{i(p+q)x}\frac{\tilde{g}_{BO}(p)\tilde{g}_{BO}(q)}{C_{BO}^2\omega_{BO}}(2\omega_{BO})[b_{BO},b^\dag_{BO}]\nonumber\\
&=&3\beta^2\int dx\ \sech^2(\beta x)\frac{{g}^2_{BO}(x)}{C_{BO}^2\omega_{BO}}b^\dag_{BO}b_{BO}\nonumber\\
&&-\frac{3\beta^2}{2}\int dx\ \sech^2(\beta x) \frac{{g}_{BO}(x)g^*_{BO}(x)}{C_{BO}^2\omega_{BO}}\nonumber\\
&&-\frac{3\beta^2}{2}\int dx\pin{p}\pin{q} \frac{\sech^2(\beta x)}{\omega_p}e^{i(p+q)x}\frac{\tilde{g}_{BO}(p)\tilde{g}_{BO}(q)}{C_{BO}^2}.
\eea
The $b^\dag b$ term cancels that in (\ref{hbo}), leaving the last two lines, which are scalars.

Summarizing, the odd bound state contribution to the Hamiltonian is
\beq
H_{BO}=\omega_{BO} b^\dag_{BO} b_{BO}+Q_{BO}
\eeq
where the scalar term $Q_{BO}$ is
\bea
Q_{BO}&=&\frac{1}{4}\frac{I_2}{C_{BO}^2\omega_{BO}}-\frac{3\beta^2}{2}\int dx\pin{p}\pin{q} \frac{\sech^2(\beta x)}{\omega_p}e^{i(p+q)x}\frac{\tilde{g}_{BO}(p)\tilde{g}_{BO}(q)}{C_{BO}^2}\nonumber\\
&&-\frac{3\beta^2}{2}\int dx\ \sech^2(\beta x) \frac{{g}_{BO}(x)g^*_{BO}(x)}{C_{BO}^2\omega_{BO}}.
\eea
Using the equations of motion one can again remove the $\sech$ from $Q_{BO}$
\bea
Q_{BO}&=&\frac{1}{4}\pin{p}\left[\frac{(\omega_p-\omega_{BO})^2}{\omega_{BO}}+\frac{\omega_p^2-\omega_{BO}^2}{\omega_p}-\frac{\omega_p^2-\omega_{BO}^2}{\omega_{BO}}
\right]\frac{\tilde{g}^2_{BO}(p)}{C_{BO}^2}\nonumber\\
&=&-\frac{1}{4}\pin{p}\frac{(\omega_p-\omega_{BO})^2}{\omega_p}\frac{\tilde{g}^2_{BO}(p)}{C_{BO}^2} . \label{qbo}
\eea
In this derivation we repeatedly used the fact that $\tilde{g}_{BO}(p)$ is odd.  Notice that our result is nearly identical to that in the continuum case (\ref{qc}), except that the integral over $k$ is gone as there is only one state.

\subsection{Even Bound State Contribution}

Recall that the even bound state is a Goldstone mode and so has zero frequency $\omega_{BE}$.  Therefore instead of oscillator modes satisfying the Heisenberg algebra we introduced zero modes $\phi_0$ and $\pi_0$ which satisfy the canonical algebra.  Their contribution to $H_0$ is
\bea
H_{BE,0}&=&\pin{p} \omega_p a^\dag_{BE,p} a_{BE,p}\label{hbe}\\
&=&\frac{1}{2}\pin{p}\frac{\tilde{g}_{BE}(p)\tilde{g}_{BE}(-p)}{C_{BE}^2}\left(\omega_p^2\phi^2_0+\pi^2_0+i\omega_p[\phi_0,\pi_0]\right)\nonumber\\
&=&\frac{\pi_0^2}{2}+\frac{1}{2}\pin{p}\frac{\tilde{g}_{BE}(p)\tilde{g}_{BE}(p)}{C_{BE}^2}\left(\omega_p^2\phi^2_0-\omega_p\right). \nonumber
\eea
The first term is the kinetic energy of the kink arising from a plane wave superposition of kinks with different centers $x_0$ and phase proportional to $x_0$ times the eigenvalue of $\pi_0$.

The contribution from $\tilde{T}_2$ is
\bea
\tilde{T}_{BE,2}&=&-\frac{3\beta^2}{2}\int dx\pin{p}\pin{q}\frac{\sech^2(\beta x)}{\sqrt{\omega_p\omega_q}}e^{i(p+q)x}\\
&&\times \left(a^\dag_{BE,p} a^\dag_{BE,q}+a^\dag_{BE,p} a_{BE,-q}+a^\dag_{BE,q}a_{BE,-p}+a_{BE,-q}a_{BE,-p}\right)\nonumber\\
&=&-\frac{3\beta^2}{4}\int dx \pin{p}\pin{q} \sech^2(\beta x)e^{i(p+q)x}\frac{\tilde{g}_{BE}(p)\tilde{g}_{BE}(q)}{C_{BE}^2}\nonumber\\
&\times&\left[4\phi_0^2+i\left(\frac{1}{\omega_p}+\frac{1}{\omega_q}\right)[\phi_0,\pi_0]\right]\nonumber\\
&=&-3\beta^2\int dx\ \sech^2(\beta x)\frac{g_{BE}(x)}{C_{BE}^2}\pin{p}{\tilde{g}}_{BE}(p)e^{-ipx}\phi_0^2\nonumber\\
&&+\frac{3\beta^2}{2}\int dx \pin{p}\pin{q} \sech^2(\beta x)e^{i(p+q)x}\frac{\tilde{g}_{BE}(p)\tilde{g}_{BE}(q)}{C_{BE}^2\omega_p}\nonumber\\
&=&A+B\nonumber
\eea
where $A$ is the term proportional to $\phi_0^2$.

Using the equations of motion 
\beq
6\beta^2\int dx\pin{p}e^{-ipx}\sech^2(\beta x) g_{BE}(x) =\pin{p}{\tilde{g}}_{BE}(-p)\omega_p^2
\eeq
one sees that $A$ cancels the $\phi_0^2$ term in (\ref{hbe}).

We are left with
\beq
H_{BE}=\frac{\pi_0^2}{2}+Q_{BE}
\eeq
where
\bea
Q_{BE}
&=&\left(\frac{1}{4}-\frac{1}{2}\right)\pin{p}\frac{\tilde{g}_{BE}(p)\tilde{g}_{BE}(p)}{C_{BE}^2}\omega_p\nonumber\\
&=&-\frac{1}{4}\pin{p}\frac{\tilde{g}_{BE}(p)\tilde{g}_{BE}(p)}{C_{BE}^2}\omega_p.\label{qbe}
\eea
It is of the same form as $Q_C$ in (\ref{qc}) and $Q_{BO}$ in (\ref{qbo}), as $\omega_{BE}=0$.

\subsection{Putting It All Together}

Now we are ready to evaluate the mass of the kink.  Classically the mass is $E_{cl}$ as given in Eq.~(\ref{ecl}).  The quantum correction $E_K-E_{cl}$ is given by the Schrodinger equation (\ref{trunc}) as the minimal eigenvalue of $H_{PT}$.  We have seen that $H_{PT}$ is
\beq
H_{PT}=\pin{k}\omega_k b^\dag_k b_k+\omega_{BO} b^\dag_{BO} b_{BO}
+\frac{\pi_0^2}{2}+Q \label{hfin}
\eeq
where Eqs.~(\ref{qc}), (\ref{qbo}) and (\ref{qbe}) give 
\bea
Q&=&Q_C+Q_{BO}+Q_{BE} \label{q}\\
&=&
-\frac{1}{4}\pin{k}\pin{p}\frac{(\omega_p-\omega_k)^2}{\omega_p}\frac{\tilde{g}^2_{k}(p)}{C_{k}^2}
-\frac{1}{4}\pin{p}\frac{(\omega_p-\omega_{BO})^2}{\omega_p}\frac{\tilde{g}^2_{BO}(p)}{C_{BO}^2}\nonumber\\
&&
-\frac{1}{4}\pin{p}\frac{\tilde{g}^2_{BE}(p)}{C_{BE}^2}\omega_p\nonumber
\eea
which is a scalar.  

The lowest energy state $\co_1|-\rangle$ is one which satisfies
\beq
b\co_1|-\rangle=\pi_0\co_1|-\rangle=0. \label{bpi}
\eeq
Of course the eigenstates of $\pi_0$ are nonnormalizable plane waves.  However normalized states exist for which the expectation value of $\pi_0^2$ is as small as desired, although strictly positive.

The form (\ref{q}) of $Q$ is in line with intuition from second order perturbation theory.    The weight $\tilde{g}^2_{k}(p)/C_k^2$ is the overlap squared of a $PT$ eigenstate $k$ and a plane wave momentum $p$ eigenstate.  Therefore at each PT state $k$, this computes the expectation value of $(\omega_p-\omega_k)^2/\omega_p$, averaged over $p$
\beq
Q\sim -\frac{1}{4}\sum_k \left\langle \frac{(\omega_p-\omega_k)^2}{\omega_p}\right\rangle_p
\eeq
where we recall that $\omega$ is energy.    The perturbation is nonvanishing because $H_{PT}$ has a potential well.  As a result, at first order in perturbation theory the state $\co_1|-\rangle$ differs from $|-\rangle$ by of order $(\omega_p-\omega_k)/\omega_p$.  This leads to a shift in energy at second order in perturbation theory of $(\omega_p-\omega_k)^2/\omega_p$.  This intuition will be tested in future work when we compute $\co_1$.

The state $|-\rangle$ is fixed, and so (\ref{bpi}) is a condition on the operator $\co_1$.  Any such state will satisfy
\beq
H_{PT}\co_1|-\rangle=Q\co_1|-\rangle.
\eeq
Therefore by (\ref{trunc}) the kink mass, which is the lowest energy of a kink, is
\beq
E_K=E_{cl}+Q.
\eeq
As expected, $Q$ is the quantum correction to the kink mass.  In the approximation (\ref{trunc}), in which $H_I$ has been dropped, it is exact.  The inclusion of $H_I$ will include corrections which are subdominant in our $\sqrt{\lambda}$ expansion.

What is $Q$?  While we have not been able to perform any of these integrals analytically, numerically we have found
\beq
Q_C=-0.082\beta \hsp
Q_{BO}=-0.040\beta \hsp
Q_{BE}=-0.544\beta\hsp
Q=-0.666\beta
\eeq
where we recall that
\beq
\beta=\frac{m}{2}=\sqrt{\frac{\lambda}{2}}v.
\eeq
$\beta$ is equal to $m/\sqrt{2}$ in the notation of Ref.~\cite{dhn2}.  Our expression for the total quantum correction to the energy agrees with theirs to the three-digit numerical precision that we have obtained, although the three individual contributions differ.

\section{Remarks} \label{remsez}

\subsection{Three Corollaries}

After a long calculation, we have arrived at the same mass found in Ref.~\cite{dhn2} via a short computation.  What have we gained?

\subsubsection{Other Solutions}

For any (1+1)-dimensional theory of a scalar $\phi$ with a canonical kinetic term and a potential $V[\phi]$, with a classical solution $f(x)$, we could have done the same calculation.  An operator $\df$ creates the solution, but leads to a new Hamiltonian $H\p$ defined by
\beq
H\df=\df H\p
\eeq
by shifting the kinetic term by
\beq
\delta \mathcal{L}=\frac{1}{2}:V^{\prime\prime}[f(x)]:\phi^2(x): \label{newpot}
\eeq
and modifying the higher order interactions.  The fact that $f(x)$ solves the classical equations of motion guarantees that $H\p$ does not contain any terms linear in $\phi$.  The new Hamiltonian could be truncated to second order to obtain a new Schrodinger equation generalizing our Eq.~(\ref{trunc}).  This new Hamiltonian would have different eigenfunctions $g(x)$ which define new operators $b$ and $b^\dag$.  However just the same steps could be followed as above to write $H\p$ as a sum of $b^\dag b$ terms and a scalar.  The $V^{\prime\prime}$ terms could be eliminated by the equation of motion, which is the same as above due to the canonical kinetic term.  

The formula Eq.~(\ref{q}) can be written as follows
\bea
Q&=&Q_C+\sum_I Q_{BI} \label{qgen}\\
Q_C&=&-\frac{1}{4}\pin{k}\pin{p}\frac{(\omega_p-\omega_k)^2}{\omega_p}\frac{\tilde{g}^2_{k}(p)}{C_{k}^2}\nonumber\\
Q_{BI}&=&-\frac{1}{4}\pin{p}\frac{(\omega_p-\omega_{BI})^2}{\omega_p}\frac{\tilde{g}^2_{BI}(p)}{C_{BI}^2}\nonumber
\eea
where the index $I$ runs over all bound states.  The procedure described above suggests that this formula yields the quantum correction to the mass of any time-independent classical solution in any such theory.

In the case treated in this paper, the functions $\tilde{g}$ contained Dirac delta functions, which occur only at $p=k$ and so do not contribute to (\ref{qgen}).  This in turn is a result of the fact that scattering in the potential $H_{PT}$ is reflectionless.  Had this not been the case, there may have been another delta function at $p=-k$.  Such a delta function would also not contribute, as the prefactor $(\omega_p-\omega_k)$ also vanishes at $p=-k$.

\subsubsection{The Spectrum}

The form (\ref{hfin}) for the Hamiltonian provides the entire spectrum.  All mass eigenstates are created by combinations of the three following actions.  First, one may boost the solution to tune $\pi_0^2$ to any positive value.  This will increase the energy by half the eigenvalue of $\pi_0^2$.  Second, one may act with an arbitrary natural number $k$ of $b_{BO}^\dag$, exciting the odd bound state.  This will increase the energy by $k\omega_{BO}$, where $\omega_{BO}$ is given in Eq.~(\ref{wbo}).  Finally one may act with any number of $b^\dag_k$.  Each increases the energy by $\omega_k$, where $\omega_k$ is given in (\ref{kdef}).   This completely characterizes the spectrum of the 1-kink sector up to $O(m)$.

\subsubsection{The Soliton Operator}

Our kink is created from the vacuum $|-\rangle$ by the operator $\df\co_1$.  While the displacement operator is defined by Eq.~(\ref{df}), we have not found a candidate operator $\co_1$.  The operator $\co_1$ is defined by
\beq
H_{PT}\co_1|-\rangle=Q\co_1|-\rangle
\eeq
which is equivalent to the two conditions (\ref{bpi}).  Again dropping interaction terms, $|-\rangle$ satisfies
\beq
a_p|-\rangle=0. \label{ameno}
\eeq

Combining Eqs.~(\ref{bpi}) and (\ref{ameno}) we find that $\co_1$ must satisfy
\beq
b_k\co_1=\co_1 A(a_p)\hsp  b_{BO}\co_1=\co_1 B(a_p)\hsp \pi_0\co_1=\co_1 C(a_p) \label{bog2}
\eeq
where $A$, $B$ and $C$ are arbitrary functions of all of the annihilation operators $a_p$.  These three necessary and sufficient conditions for $\co_1$ are reminiscent of those for a Bogoliubov transform from the $b$ to the $a$, suggesting that $\co_1$ is a kind of squeeze operator.  We will try to solve (\ref{bog2}) in future work and compare our answer with the operator found in Ref.~\cite{blasone}.

\subsection{What Next?}

Why are we interested in scalar field theories in 1+1 dimensions?  Beyond the horizon, our goal is the monopole in Yang-Mills theory.  There is no classical monopole solution, and so if the 't Hooft-Mandelstam mechanism for confinement \cite{thooftconf,mandconf} is to be realized, the monopole will be an operator.  We would like to find such an operator and use Hamiltonian methods such as those used here to show that it is tachyonic.  

However we are not strong enough to guess this operator from scratch.  We need to guess a good Ansatz, and for this we will first try to solve the corresponding problem in $\mathcal{N}=2$ SQCD \cite{sw2}, where the monopole becomes tachyonic after a soft breaking to $\mathcal{N}=1$.  This case is similar to QCD in that it is strongly coupled and there is no semiclassical monopole.  In the $\mathcal{N}=2$ case there is a Higgs field, unlike QCD, however its VEV is small and the monopole mass is instead dominated by instantons.  Here the monopole which condenses is connected to a semiclassical monopole by a continuous deformation of the theory in which one turns on a bare mass for the hypermultiplets.   In the massive case, the theory is weakly coupled in the infrared and so we can find the monopole operator using perturbation theory, as is done for the kink here.  Then the key step will be to follow it through the deformation to strong coupling.  This may be possible because the monopole is BPS, and so the equations to be solved are first order and not second order like (\ref{trunc}).  Nonetheless, we will need to solve these equations exactly, not perturbatively, to obtain the monopole operator in the regime where it condenses.

To prepare ourselves for this exact calculation, we wish to do the same with the kink.  The $\phi^4$ kink studied in this paper also exists in supersymmetric field theories, where it may be BPS.  Therefore, before moving to gauge theories in (3+1)-dimensions, we wish to try this program on the operator which creates the BPS kink.  We intend to first construct it using perturbation theory, and then attempt to use the BPS equations to follow it to strong coupling.  If we cannot succeed with the kink operator, it is unlikely that we may succeed with the monopole operator.  

\appendix

\section{Hypergeometric Functions}

The solutions of Eqs.~(\ref{fkeq}) are known in terms of the ordinary hypergeometric functions ${}_2F_1$
\beq
F\left(\frac{3+ik}{2},\frac{3-ik}{2};\frac{1}{2};-{\rm{sinh}}^2(x)\right)
{\rm{\ \ \ and\ \ \ }}
F\left(\frac{4+ik}{2},\frac{4-ik}{2};\frac{3}{2};-{\rm{sinh}}^2(x)\right). \label{nof}
\eeq
We leave the subscripts implicit as all hypergeometric functions will be ordinary.  Our first goal is to compute these functions.

Mathematica is able to calculate a simpler function
\beq
F\left(a,1-a;\frac{1}{2};{\rm{sin}}^2(z)\right)=\frac{\cos((2a-1)z)}{\cos(z)}. \label{maf}
\eeq
To go from (\ref{maf}) to (\ref{nof}) we will need an analytic continuation and also Gauss' contiguous relations, which allow one to shift the first three arguments by arbitrary integers.

To derive Gauss' contiguous relations, one uses the definition of the hypergeometric functions
\beq
F\left(a,b;c;y\right)=\sum_{n=0}^\infty \frac{(a)_n(b)_n}{(c)_n}\frac{y^n}{n!} \label{hfdef}
\eeq
where we have used the rising Pochhammer symbol
\beq
(q)_n=q(q+1)\cdots (q+n-1).
\eeq
The derivative of (\ref{hfdef}) is readily computed
\bea
\partial_y F\left(a,b;c;y\right)
&=&\sum_{n=1}^\infty \frac{(a)_n(b)_n}{(c)_n}\frac{y^{n-1}}{(n-1)!}
=\frac{ab}{c}\sum_{n=1}^\infty \frac{(a+1)_{n-1}(b+1)_{n-1}}{(c+1)_{n-1}}\frac{y^{n-1}}{(n-1)!}
\nonumber\\
&=&\frac{ab}{c}F\left(a+1,b+1;c+1;y\right).
\eea
Similarly
\bea
y\partial_y F\left(a,b;c;y\right)
&=&\sum_{n=1}^\infty n \frac{(a)_n(b)_n}{(c)_n}\frac{y^{n}}{n!}=\sum_{n=1}^\infty (b+n-b) \frac{(a)_n(b)_n}{(c)_n}\frac{y^{n}}{n!}\nonumber\\
&=&\sum_{n=0}^\infty  \frac{(a)_{n}(b)_{n+1}}{(c)_n}\frac{y^{n}}{n!}-b\sum_{n=0}^\infty \frac{(a)_n(b)_n}{(c)_n}\frac{y^n}{n!} \nonumber\\
&=&b\left(F\left(a,b+1;c;y\right)-F\left(a,b;c;y\right)\right).
\eea
Thus one may increase the arguments of the hypergeometric functions using the identities
\bea
F\left(a,b+1;c;y\right)&=&F\left(a,b;c;y\right)+\frac{1}{b}y\partial_y F\left(a,b;c;y\right) \label{bpiu}\\
F\left(a+1,b;c;y\right)&=&F\left(a,b;c;y\right)+\frac{1}{a}y\partial_y F\left(a,b;c;y\right) \label{apiu}\\
F\left(a+1,b+1;c+1;y\right)&=&\frac{c}{ab}\partial_y F\left(a,b;c;y\right). \label{tpiu}
\eea
The second identity falls from the first, using the $a\leftrightarrow b$ symmetry of (\ref{hfdef}).

We begin our recursion with (\ref{apiu}) reexpressed in terms of $y$
\beq
F\left(a,1-a;\frac{1}{2};y\right)=\frac{\cos((2a-1){\asin(\sqrt{y})})}{\sqrt{1-y}}. 
\eeq
Applying the identity (\ref{apiu}) yields
\bea
F\left(a+1,1-a;\frac{1}{2};y\right)&=&
-\frac{\sqrt{y}(2a-1)}{2a(1-y)}\sin((2a-1){\asin(\sqrt{y})})\\
&&+\left(\frac{1}{\sqrt{1-y}}+\frac{y}{2a(1-y)^{3/2}}\right)\cos((2a-1){\asin(\sqrt{y})}). \nonumber
\eea
We now make the replacement $a\rightarrow a-1$ to obtain
\bea
F\left(a,2-a;\frac{1}{2};y\right)&=&
-\left(\frac{2a-3}{2a-2}\right)\frac{\sqrt{y}}{1-y}\sin((2a-3){\asin(\sqrt{y})})\label{stella}\\
&&\frac{(2a-2)-(2a-3)y}{(2a-2)(1-y)^{3/2}}\cos((2a-3){\asin(\sqrt{y})}).\nonumber
\eea
Next we apply (\ref{bpiu}) to obtain the first desired hypergeometric function
\bea
F\left(a,3-a;\frac{1}{2};y\right)&=&
\frac{3(2a-3)\sqrt{y}}{(2a-4)(2a-2)(1-y)^2}\sin((2a-3){\asin(\sqrt{y})})\\
&&\left(\frac{1}{(1-y)^{3/2}}-\frac{3y}{(2a-4)(2a-2)(1-y)^{5/2}} \right)\cos((2a-3){\asin(\sqrt{y})}).\nonumber
\eea
Choosing
\beq
a=\frac{3+ik}{2}\hsp y=-{\rm{sinh}}^2(x)
\eeq
one finds
\bea
F\left(\frac{3+ik}{2},\frac{3-ik}{2};\frac{1}{2};-{\rm{sinh}}^2(x)\right)&=&-\frac{3k}{1+k^2}\frac{\sinh(x)}{\cosh^4(x)}\sin(kx)\\
&+&\left(\frac{1}{\cosh^3(x)}-\frac{3}{1+k^2}\frac{\sinh^2(x)}{\cosh^5(x)}\right)\cos(kx).\nonumber
\eea

To find the other needed function, one applies the identities
\bea
\sin((a-1)x)&=&-\cos(ax)\sin(x)+\sin(ax)\cos(x)\nonumber\\
\cos((a-1)x)&=&\cos(ax)\cos(x)+\sin(ax)\sin(x)
\eea
to (\ref{stella}) yielding
\bea
F\left(a,2-a;\frac{1}{2};y\right)&=&
\frac{\sqrt{y}}{(2a-2)(1-y)^{3/2}}\sin((2a-2){\asin(\sqrt{y})})\label{stella}\\
&&\frac{1}{1-y}\cos((2a-2){\asin(\sqrt{y})}).\nonumber
\eea
Then the identity (\ref{tpiu}) gives the general form of the desired function
\bea
F\left(a+1,3-a;\frac{3}{2};y\right)&=&
\frac{(2a-1)(2a-3)+3y}{2a(2a-2)(2a-4)\sqrt{y}(1-y)^{5/2}}\sin((2a-2){\asin(\sqrt{y})})\nonumber\\
&&-\frac{3}{2a(2a-4)(1-y)^2}\cos((2a-2){\asin(\sqrt{y})})\label{stella}
\eea
and so in particular
\bea
F\left(\frac{4+ik}{2},\frac{4-ik}{2};\frac{3}{2};-{\rm{sinh}}^2(x)\right)&=&
\left(\frac{k^2+1-3{\rm{tanh}}^2(x)}{k(k^2+4)\sinh(x)\cosh^3(x)}\right)\sin(kx)\nonumber\\
&&+\frac{3}{(k^2+4)\cosh^4(x)} \cos(kx). \label{hyp}
\eea

\section* {Acknowledgement}

\noindent
JE is supported by the CAS Key Research Program of Frontier Sciences grant QYZDY-SSW-SLH006 and the NSFC MianShang grants 11875296 and 11675223.   JE also thanks the Recruitment Program of High-end Foreign Experts for support.


\begin{thebibliography}{99}

\bibitem{gervais}
  J.~L.~Gervais and B.~Sakita,
  ``Extended Particles in Quantum Field Theories,''
  Phys.\ Rev.\ D {\bf 11} (1975) 2943.
  doi:10.1103/PhysRevD.11.2943

\bibitem{delfino}
  G.~Delfino, W.~Selke and A.~Squarcini,
  ``Vortex mass in the three-dimensional $O(2)$ scalar theory,''
  Phys.\ Rev.\ Lett.\  {\bf 122} (2019) no.5,  050602
  doi:10.1103/PhysRevLett.122.050602
  [arXiv:1808.09276 [cond-mat.stat-mech]].

\bibitem{davies}
  D.~Davies,
  ``Quantum Solitons in any Dimension: Derrick's Theorem v. AQFT,''
  arXiv:1907.10616 [hep-th].

\bibitem{hepp}
  K.~Hepp,
  ``The Classical Limit for Quantum Mechanical Correlation Functions,''
  Commun.\ Math.\ Phys.\  {\bf 35} (1974) 265.
  doi:10.1007/BF01646348

\bibitem{mandelstamkink}
  S.~Mandelstam,
  ``Soliton Operators for the Quantized Sine-Gordon Equation,''
  Phys.\ Rev.\ D {\bf 11} (1975) 3026.
  doi:10.1103/PhysRevD.11.3026

\bibitem{taylor78}
  J.~G.~Taylor,
  ``Solitons as Infinite Constituent Bound States,''
  Annals Phys.\  {\bf 115} (1978) 153.
  doi:10.1016/0003-4916(78)90179-3

\bibitem{dhn2}
  R.~F.~Dashen, B.~Hasslacher and A.~Neveu,
  ``Nonperturbative Methods and Extended Hadron Models in Field Theory 2. Two-Dimensional Models and Extended Hadrons,''
  Phys.\ Rev.\ D {\bf 10} (1974) 4130.
  doi:10.1103/PhysRevD.10.4130


\bibitem{flugge}
S. Fl\"ugge,
``Practical Quantum Mechanics,"
Springer-Verlag Berlin Heidelberg (1999),
doi:10.1007/978-3-642-61995-3


\bibitem{lekner}
J. Lekner,
``Reflectionless eigenstates of the sech${}^2$ potential,"
Am. J. Phys. 75 (2007) 1151,
doi:10.1119/1.278701

\bibitem{blasone}
  M.~Blasone and P.~Jizba,
  ``Topological defects as inhomogeneous condensates in quantum field theory: Kinks in (1+1)-dimensional lambda psi**4 theory,''
  Annals Phys.\  {\bf 295} (2002) 230
  doi:10.1006/aphy.2001.6215
  [hep-th/0108177].

\bibitem{thooftconf}
  G.~'t Hooft,
  ``Topology of the Gauge Condition and New Confinement Phases in Nonabelian Gauge Theories,''
  Nucl.\ Phys.\ B {\bf 190} (1981) 455.
  doi:10.1016/0550-3213(81)90442-9

\bibitem{mandconf}
  S.~Mandelstam,
  ``Vortices and Quark Confinement in Nonabelian Gauge Theories,''
  Phys.\ Rept.\  {\bf 23} (1976) 245.
  doi:10.1016/0370-1573(76)90043-0


\bibitem{sw2}
  N.~Seiberg and E.~Witten,
  ``Electric - magnetic duality, monopole condensation, and confinement in N=2 supersymmetric Yang-Mills theory,''
  Nucl.\ Phys.\ B {\bf 426} (1994) 19
   Erratum: [Nucl.\ Phys.\ B {\bf 430} (1994) 485]
  doi:10.1016/0550-3213(94)90124-4, 10.1016/0550-3213(94)00449-8
  [hep-th/9407087].

\end{thebibliography}

\begin{thebibliography}{99}

\bibitem{naumov1}
  D.~V.~Naumov and V.~A.~Naumov,
  ``A Diagrammatic treatment of neutrino oscillations,''
  J.\ Phys.\ G {\bf 37} (2010) 105014
  doi:10.1088/0954-3899/37/10/105014
  [arXiv:1008.0306 [hep-ph]].

\bibitem{naumov2}
  D.~V.~Naumov,
  ``On the Theory of Wave Packets,''
  Phys.\ Part.\ Nucl.\ Lett.\  {\bf 10} (2013) 642
  doi:10.1134/S1547477113070145
  [arXiv:1309.1717 [quant-ph]].

\bibitem{dayadec}
  F.~P.~An {\it et al.} [Daya Bay Collaboration],
  ``Study of the wave packet treatment of neutrino oscillation at Daya Bay,''
  Eur.\ Phys.\ J.\ C {\bf 77} (2017) no.9,  606
  doi:10.1140/epjc/s10052-017-4970-y
  [arXiv:1608.01661 [hep-ex]].

\bibitem{t2k}
  K.~Abe {\it et al.} [T2K Collaboration],
  ``Search for CP Violation in Neutrino and Antineutrino Oscillations by the T2K Experiment with $2.2\times10^{21}$ Protons on Target,''
  Phys.\ Rev.\ Lett.\  {\bf 121} (2018) no.17,  171802
  doi:10.1103/PhysRevLett.121.171802
  [arXiv:1807.07891 [hep-ex]].


\bibitem{steven}
  Y.~L.~Chan, M.-C.~Chu, K.~M.~Tsui, C.~F.~Wong and J.~Xu,
  ``Wave-packet treatment of reactor neutrino oscillation experiments and its implications on determining the neutrino mass hierarchy,''
  Eur.\ Phys.\ J.\ C {\bf 76} (2016) no.6,  310
  doi:10.1140/epjc/s10052-016-4143-4
  [arXiv:1507.06421 [hep-ph]].



\bibitem{beuthe}
  M.~Beuthe,
  ``Oscillations of neutrinos and mesons in quantum field theory,''
  Phys.\ Rept.\  {\bf 375} (2003) 105
  doi:10.1016/S0370-1573(02)00538-0
  [hep-ph/0109119].

\bibitem{giunti2012}
 C.~Giunti,
  ``Neutrino wave packets in quantum field theory,''
  JHEP {\bf 0211} (2002) 017
  doi:10.1088/1126-6708/2002/11/017
  [hep-ph/0205014].

\bibitem{noi1}
  J.~Evslin, H.~Mohammed, E.~Ciuffoli and Y.~Zhou,
  ``Entangled Neutrino States in a Toy Model QFT,''
  Eur.  Phys. J. C In Press,
  arXiv:1902.03934 [hep-ph].

\bibitem{xin}
  X~Qian and J.~C.~Peng,
  ``Physics with Reactor Neutrinos,''
  Rept.\ Prog.\ Phys.\  {\bf 82} (2019) no.3,  036201
  doi:10.1088/1361-6633/aae881
  [arXiv:1801.05386 [hep-ex]].

\end{thebibliography}

\begin{thebibliography}{99}

\bibitem{boya2011}
  D.~Boyanovsky,
  ``Short baseline neutrino oscillations: when entanglement suppresses coherence,''
  Phys.\ Rev.\ D {\bf 84} (2011) 065001
  doi:10.1103/PhysRevD.84.065001
  [arXiv:1106.6248 [hep-ph]].


\bibitem{accdec}
  B.~J.~P.~Jones,
  ``Dynamical pion collapse and the coherence of conventional neutrino beams,''
  Phys.\ Rev.\ D {\bf 91} (2015) no.5,  053002
  doi:10.1103/PhysRevD.91.053002
  [arXiv:1412.2264 [hep-ph]].

\bibitem{steven}
  Y.~L.~Chan, M.-C.~Chu, K.~M.~Tsui, C.~F.~Wong and J.~Xu,
  ``Wave-packet treatment of reactor neutrino oscillation experiments and its implications on determining the neutrino mass hierarchy,''
  Eur.\ Phys.\ J.\ C {\bf 76} (2016) no.6,  310
  doi:10.1140/epjc/s10052-016-4143-4
  [arXiv:1507.06421 [hep-ph]].


\bibitem{dayadec}
  F.~P.~An {\it et al.} [Daya Bay Collaboration],
  ``Study of the wave packet treatment of neutrino oscillation at Daya Bay,''
  Eur.\ Phys.\ J.\ C {\bf 77} (2017) no.9,  606
  doi:10.1140/epjc/s10052-017-4970-y
  [arXiv:1608.01661 [hep-ex]].


\bibitem{revival}
  K.~Kiers and N.~Weiss,
  ``Neutrino oscillations in a model with a source and detector,''
  Phys.\ Rev.\ D {\bf 57} (1998) 3091
  doi:10.1103/PhysRevD.57.3091
  [hep-ph/9710289].

\bibitem{nuss76}
  S.~Nussinov,
  ``Solar Neutrinos and Neutrino Mixing,''
  Phys.\ Lett.\  {\bf 63B} (1976) 201.
  doi:10.1016/0370-2693(76)90648-1

\bibitem{wilczek}
  L.~Krauss and F.~Wilczek,
  ``Solar Neutrino Oscillations,''
  Phys.\ Rev.\ Lett.\  {\bf 55} (1985) 122.
  doi:10.1103/PhysRevLett.55.122

\bibitem{rich}
 J.~Rich,
  ``The Quantum mechanics of neutrino oscillations,''
  Phys.\ Rev.\ D {\bf 48} (1993) 4318.
  doi:10.1103/PhysRevD.48.4318

\bibitem{boriserr}
  B.~Kayser and J.~Kopp,
  ``Testing the wave packet approach to neutrino oscillations in future experiments,''
  arXiv:1005.4081 [hep-ph].

\bibitem{giunti2012}
 C.~Giunti,
  ``Neutrino wave packets in quantum field theory,''
  JHEP {\bf 0211} (2002) 017
  doi:10.1088/1126-6708/2002/11/017
  [hep-ph/0205014].

\bibitem{zurek}
  W.~H.~Zurek,
  ``Environment induced superselection rules,''
  Phys.\ Rev.\ D {\bf 26} (1982) 1862.
  doi:10.1103/PhysRevD.26.1862

\bibitem{mcgreevy}
A.~Kobach, A.~V.~Manohar and J.~McGreevy,
  ``Neutrino Oscillation Measurements Computed in Quantum Field Theory,''
  Phys.\ Lett.\ B {\bf 783} (2018) 59
  doi:10.1016/j.physletb.2018.06.021
  [arXiv:1711.07491 [hep-ph]].

\bibitem{cgl}
  A.~G.~Cohen, S.~L.~Glashow and Z.~Ligeti,
  ``Disentangling Neutrino Oscillations,''
  Phys.\ Lett.\ B {\bf 678} (2009) 191
  doi:10.1016/j.physletb.2009.06.020
  [arXiv:0810.4602 [hep-ph]].

\bibitem{mcdonald}
``Oscillations and decoherence,"
Kirk T McDonald, Talk at NuFact 2013,
August 23, 2013, Beijing, China.

\bibitem{naumov1}
  D.~V.~Naumov and V.~A.~Naumov,
  ``A Diagrammatic treatment of neutrino oscillations,''
  J.\ Phys.\ G {\bf 37} (2010) 105014
  doi:10.1088/0954-3899/37/10/105014
  [arXiv:1008.0306 [hep-ph]].


\bibitem{naumov2}
  D.~V.~Naumov,
  ``On the Theory of Wave Packets,''
  Phys.\ Part.\ Nucl.\ Lett.\  {\bf 10} (2013) 642
  doi:10.1134/S1547477113070145
  [arXiv:1309.1717 [quant-ph]].


\bibitem{akqft}
  E.~K.~Akhmedov and A.~Y.~Smirnov,
  ``Neutrino oscillations: Entanglement, energy-momentum conservation and QFT,''
  Found.\ Phys.\  {\bf 41} (2011) 1279
  doi:10.1007/s10701-011-9545-4
  [arXiv:1008.2077 [hep-ph]].

\bibitem{akmoss}
  E.~K.~Akhmedov, J.~Kopp and M.~Lindner,
  ``Oscillations of Mossbauer neutrinos,''
  JHEP {\bf 0805} (2008) 005
  doi:10.1088/1126-6708/2008/05/005
  [arXiv:0802.2513 [hep-ph]].


\bibitem{colemanlect}
 B.~G.~G.~Chen, D.~Derbes, D.~Griffiths, B.~Hill, R.~Sohn and Y.~S.~Ting,
  ``Lectures of Sidney Coleman on Quantum Field Theory,''
  doi:10.1142/9371 .


\bibitem{review}
  M.~Beuthe,
  ``Oscillations of neutrinos and mesons in quantum field theory,''
  Phys.\ Rept.\  {\bf 375} (2003) 105
  doi:10.1016/S0370-1573(02)00538-0
  [hep-ph/0109119].



\end{thebibliography}
\end{document}